\documentclass{article}
\usepackage{arxiv}
\usepackage{color}
\usepackage{amsmath}
\usepackage{amsfonts}
\usepackage{bm}
\usepackage{soul}
\usepackage{graphicx}
\usepackage{hyperref}
\usepackage{authblk}

\usepackage[natbibapa]{apacite}
\let\svthefootnote\thefootnote
\newcommand\freefootnote[1]{%
  \let\thefootnote\relax%
  \footnotetext{#1}%
  \let\thefootnote\svthefootnote%
}

\DeclareMathOperator*{\argmax}{argmax}
\DeclareMathOperator*{\argmin}{argmin}

\title{Confidence is detection-like in high-dimensional spaces}
\author[1]{Wiktoria Kozyra}
\author[2,3]{Kevin O'Neill}
\author[2,3,4]{Stephen M. Fleming}
\affil[1]{Department of Experimental Psychology, University of Oxford, Oxford, UK}
\affil[2]{Department of Experimental Psychology, University College London, London, UK}
\affil[3]{Institute of Cognitive Neuroscience, University College London, London, UK}
\affil[4]{Max Planck UCL Centre for Computational Psychiatry and Ageing Research, University College London, London, UK}

\begin{document}
\maketitle

\begin{abstract}
    Confidence estimates are often ``detection-like''---driven by positive evidence in favour of a decision. This empirical observation has been interpreted as showing that human metacognition is limited by biases or heuristics. Here, we show that Bayesian confidence estimates also exhibit heightened sensitivity to decision-congruent evidence in higher-dimensional signal detection theoretic spaces, leading to detection-like confidence criteria. This effect is due to a nonlinearity induced by normalisation of confidence by a large number of unchosen alternatives. Our analysis suggests that detection-like confidence is rational when participants consider a greater number of hypotheses than assumed by the experimenter. Further, we show that a similar dimensionality-driven mechanism can give rise to and modulate the strength of the positive evidence biases in convolutional neural networks, linking our signal detection theoretic analysis to confidence behaviour in artificial systems.
\end{abstract}

\freefootnote{
Please cite the published version of this article:

Kozyra, W., O'Neill, K., \& Fleming, S.M. (2026). Confidence is detection-like in high-dimensional spaces. \emph{Open Mind}.

This work was supported by a UKRI/EPSRC Programme Grant [EP/V000748/1]. SMF is a CIFAR Fellow in the Brain, Mind \& Consciousness Program, and supported by UK Research and Innovation (UKRI) under the UK government's Horizon Europe funding guarantee (selected as ERC Consolidator, grant number 101043666). Parts of this work were previously presented at the Cognitive Computational Neuroscience Meeting, Oxford, August 2023.
}

\section{Introduction}
Understanding how confidence is formed in human perceptual decision-making is important for identifying sources of metacognitive failure. One prominent metacognitive bias is termed the ``positive evidence bias'' (PEB). A PEB is the empirical observation that confidence ratings (but not first-order decisions) are often insensitive to evidence against (relative to evidence for) one's choice \citep{Zylberberg2012, Peters2017a, Koizumi2015, Samaha2019, Odegaard2018}. A consequence of a PEB is that observers may become inappropriately confident when the total evidence supporting any decision option increases, even if their choice accuracy remains unaltered \citep{Rollwage2020}. 

\begin{figure}[htbp]
    \centering
    \includegraphics[width=.75\linewidth]{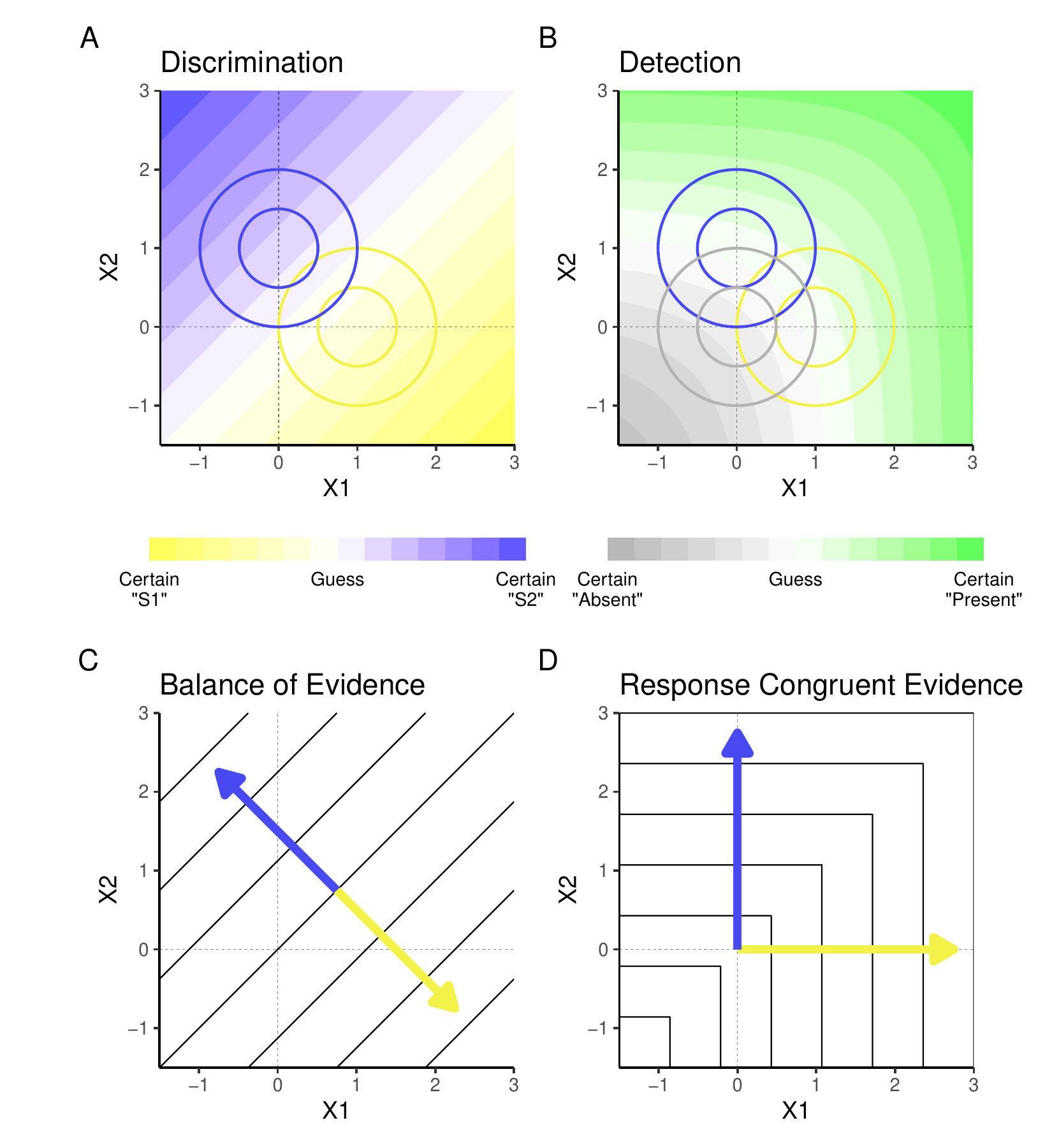}
    \caption{
    Relationship between sensory evidence and confidence for (A) discrimination tasks and (B) detection tasks in a 2-dimensional signal detection theory (SDT) model. Circles indicate bivariate Gaussian evidence distributions produced by stimulus S1 (blue), stimulus S2 (yellow), and an absent stimulus (grey). Shaded areas indicate the optimal confidence ratings for an ideal Bayesian observer. (C-D) Two heuristic strategies for responding in discrimination tasks. Under the balance of evidence heuristic (C), confidence is sensitive to relative evidence (the distance from the diagonal). Under the response congruent evidence heuristic (D), confidence is only sensitive to evidence for the chosen option (the distance along the horizontal or vertical axis). Although the balance of evidence heuristic is the optimal strategy for discrimination tasks, confidence ratings tend to resemble the response congruent evidence heuristic, which approximates the optimal strategy in detection tasks. Modified from \citet{Mazor2023}.} 
    \label{fig1}
\end{figure}

A PEB can be accounted for within signal detection theory (SDT) as confidence criteria becoming ``detection-like'' \citep{Rausch2018, Mazor2023, Hautus2021}. In SDT, the observer receives a noisy evidence signal dependent on the target stimulus and makes a choice by evaluating a set of hypotheses with respect to the evidence. For a discrimination decision between two symmetrical hypotheses, the total amount of evidence should be irrelevant to one's confidence in the decision---instead, what matters is the relative evidence supporting one decision option over the other (Figure \ref{fig1}A). For a two-dimensional SDT problem, this yields optimal decision and confidence criteria that track the balance of evidence in favour of one stimulus over the other (the distance to the main diagonal), irrespective of the total evidence (the distance along the main diagonal; Figure \ref{fig1}C). In a detection task, in contrast, the optimal decision and confidence criteria become perpendicular to the evidence axes---because what matters for detecting (rather than discriminating) a stimulus is the extent to which the amount of evidence for the dominating option deviates from the distribution of percepts expected under the \textit{absence} of any stimuli (Figure \ref{fig1}B).

Previous efforts to explain a PEB have focused on explaining why confidence criteria become detection-like. These explanations include proposals that confidence estimates are based on response-congruent evidence \citep[Figure \ref{fig1}D;][]{Maniscalco2016, Miyoshi2020}; that neural computations of confidence are preferentially informed by a subset of less normalised, ``detection-like'' neurons \citep{Maniscalco2021}; or that there is an additional contribution of stimulus visibility to confidence estimates \citep{Rausch2018}. Other studies have suggested that a PEB may arise due to a violation in the SDT assumption of equal-variance bivariate Gaussian distributions with naturalistic stimuli, which tend to show more variance along target vs. non-target dimensions \citep{Miyoshi2020}. More broadly, the PEB has typically been considered a bias or suboptimality that plagues human metacognition \citep{Koriat1980, Maniscalco2016, Miyoshi2020}.

Against this backdrop, \citet{Webb2023} showed that a PEB can emerge naturally in convolutional neural networks trained to optimise the accuracy of both choices and confidence estimates when discriminating between handwritten digits. These intriguing results indicated that the PEB may be a result of the adaptive consequences of learning the statistics of natural stimuli, rather than as a suboptimality particular to human metacognition. 

\citet{Webb2023}'s findings are important in presenting the possibility that apparent suboptimalities in human confidence may be explained by optimal computations applied to more naturalistic evidence spaces. Their results were obtained as a by-product of training a neural network on higher-dimensional stimulus sets. However, if this hypothesis is correct, it should be possible to expose such decision rules directly within the simpler framework of signal detection theory (SDT), rather than relying on a neural network to discover their emergence. In this theoretical note, we propose that at least some forms of a PEB observed in both humans and neural networks may reflect optimal confidence computation under the assumption of multidimensional evidence spaces. This framework provides a principled explanation for the emergence of detection-like confidence and generates the testable prediction that manipulating representational dimensionality should systematically modulate the strength of the positive evidence bias.

Dimensionality here refers to the number of possible features tracked by the observer---independent hypotheses for which one accumulates evidence. Classical SDT has one feature dimension encoding a property such as brightness or familiarity. 2D SDT allows two feature dimensions which are typically treated as encoding ``evidence'' for each of two distinct stimuli, such as a right-tilted or left-tilted grating \citep{Peters2015, Hautus2021}. But even when tasks force participants to classify stimuli into one of two categories defined by the experimenter, it is often natural for participants to tentatively consider a wide range of hypotheses about the stimulus before responding. For instance, in an orientation discrimination task, observers are unlikely to restrict their perceptual inferences to whether the orientation is left- or right-tilted; they may also be implicitly sensitive to other incidental stimulus properties, such as phase, spatial frequency, contrast, size, duration, and so on. To account for this possibility, here we allow for greater numbers of feature dimensions such that the evidence space becomes $k$-dimensional. We simulate a typical PEB experiment where subjects are asked to make a 2-choice discrimination between two stimulus categories, $s_1$ and $s_2$. We then compute the impact of additional ``distractor'' stimulus dimensions representing evidence for alternative stimulus categories $s_{3:k}$ on the optimal confidence criteria for two-choice discrimination. To preempt our results, we show that confidence criteria are discrimination-like when the hypothesis space is low-dimensional, but become detection-like when the space is higher-dimensional. The appearance of detection-like confidence criteria in higher-dimensional SDT spaces indicates that a PEB can be rational if subjects are computing confidence in a higher-dimensional evidence space than assumed by the experimenter. 

Finally, this prediction also provides a way to reinterpret recent findings in artificial systems. \citet{Webb2023} showed that convolutional neural networks trained to jointly optimise perceptual decisions and confidence exhibit a positive evidence bias when discriminating between handwritten digits. From the perspective developed here, this effect need not be specific to either humans or neural networks per se, but may instead reflect an incidental property of these models---namely, that confidence is computed over a high-dimensional space of alternative hypotheses. While such networks are not a direct implementation of the SDT framework developed in this paper, they nonetheless offer a useful illustrative case in which the consequences of operating over higher-dimensional representations can be examined. Accordingly, we adapt the neural network models introduced by \citep{Webb2023} to explore whether variations in representational dimensionality are accompanied by corresponding changes in the strength of detection-like confidence. Taken together, the SDT and neural network results motivate an interpretation of the positive evidence bias as an efficient heuristic that is computationally optimal under the assumption of multidimensional evidence, offering a principled account of why such a bias might arise in biological systems rather than reflecting a mere inefficiency of confidence computation.

\section{Methods}
\subsection{Signal detection theory in higher-dimensional spaces}
\subsubsection{Bivariate case}
We start with a standard 2-dimensional bivariate Gaussian SDT model (Figure \ref{fig1}). In this model, the observer is asked to determine the identity of an unknown stimulus $s \in \{s_1, s_2\}$ on the basis of a 2-dimensional sensory evidence signal $\bm{x} = [x_1, x_2]$. Each dimension of the evidence corresponds to a possible stimulus category such that $x_1$ is the evidence for $s = s_1$ and $x_2$ is the evidence for $s = s_2$. The evidence along each dimension is assumed to be independently normally distributed such that the average evidence is $\mu_\textrm{target}$ along the target dimension (representing the true stimulus category) and 0 along the other dimension:
\begin{align*}
    &x_1 \sim \begin{cases}
        \mathcal{N}\left(\mu_\textrm{target}, 1\right) & \textrm{if } s = s_1 \\
        \mathcal{N}\left(0, 1\right) & \textrm{otherwise}
    \end{cases}
    &x_2 \sim \begin{cases}
        \mathcal{N}\left(\mu_\textrm{target}, 1\right) & \textrm{if } s = s_2 \\
        \mathcal{N}\left(0, 1\right) & \textrm{otherwise}
    \end{cases}
\end{align*}
We assume that the observer infers the stimulus category $s$ by computing the posterior probabilities $P(s_1\vert\bm{x})$ and $P(s_2\vert\bm{x})$. Then, the observer chooses an action $a \in \{1,2\}$ reflecting the most probable stimulus with confidence $c \in [0, 1]$ equal to the posterior probability of the chosen option $s_a$:
\begin{align*}
    a &= \begin{cases}
        1 & \text{if }P(s_1\vert\bm{x}) > P(s_2\vert\bm{x}) \\
        2 & \text{otherwise}
    \end{cases} \\
    c &= P(s_a \vert \bm{x})
\end{align*}
Under these conditions, it is well known that this optimal strategy leads to the confidence criteria in Figure \ref{fig1}A, which can be efficiently computed using a ``balance of evidence'' heuristic \citep[Figure \ref{fig1}C; e.g.,][]{Maniscalco2016}. That is, where $f: \mathbb{R} \rightarrow [0,1]$ is a monotonic transformation,
\begin{align*}
    c  = P(s_a\vert\bm{x}) \approx f(\lvert x_1 - x_2 \rvert)
\end{align*}
The above formulation makes clear that, in the 2-dimensional setting, the optimal strategy is one that weights $x_1$ and $x_2$ equally against each other, regardless of which option is chosen.

This discrimination-like pattern of confidence (Figure \ref{fig1}A) stands in contrast to a detection-like pattern (Figure \ref{fig1}B). In a detection task, the goal is not to discriminate between possible stimuli $s_1$ and $s_2$, but to determine whether any stimulus was presented in the first place (as compared to nothing). Specifically, where $p \in \{p_0, p_1\}$ indicates stimulus absence/presence, the observer is presented with the sensory evidence $\bm{x}$ generated from the following distribution:
\begin{align*}
    &x_1 \sim \begin{cases}
        \mathcal{N}\left(\mu_\textrm{target}, 1\right) & \textrm{if } p=p_1 \textrm{ and } s = s_1 \\
        \mathcal{N}\left(0, 1\right) & \textrm{otherwise}
    \end{cases}
    &x_2 \sim \begin{cases}
        \mathcal{N}\left(\mu_\textrm{target}, 1\right) & \textrm{if } p=p_1 \textrm{ and } s = s_2 \\
        \mathcal{N}\left(0, 1\right) & \textrm{otherwise}
    \end{cases}
\end{align*}
Then, the observer selects an action indicating the presence or absence of the stimulus:
\begin{align*}
    a &= \begin{cases}
        0 & \text{if }P(p_0\vert\bm{x}) > P(p_1\vert\bm{x}) \\
        1 & \text{otherwise}
    \end{cases} \\
    c &= P(p_a \vert \bm{x})
\end{align*}
In this case, since the objective is to determine whether a stimulus was presented (ignoring which stimulus might have been presented), the optimal strategy is for confidence to increase with evidence for either stimulus \citep{Mazor2023}. However, since the strongest piece of evidence has a greater impact on accuracy, the optimal strategy is now to place confidence criteria that are roughly perpendicular to this axis (Figure \ref{fig1}B). In the strongest form of this strategy, the ``response congruent evidence'' heuristic approximates confidence using only the stronger piece of evidence (Figure \ref{fig1}D):
\begin{align*}
    c = P(p_a \vert \bm{x}) \approx f(\textrm{max}(x_1, x_2))
\end{align*}

What we aim to explain is that, in a two-alternative discrimination task, participants' confidence ratings often more closely resemble the optimal strategy for detection (rather than discrimination) tasks, even though both strategies can be approximated as simple transformations of the evidence. 

\subsubsection{Multivariate case}
We augment the above SDT model for discrimination by assuming that, although the observer is still forced to choose a stimulus among the first two alternatives ($a \in \{1, 2\}$ representing the observer's choice that $s=s_1$ and $s=s_2$), they initially entertain a larger hypothesis space of $k$ stimuli ($s \in \{s_1 \ldots s_k\}$). As before, we assume that the observer encodes the sensory evidence $\bm{x} = [x_1 \ldots x_k]$, with $x_i$ representing the evidence for the corresponding stimulus category $s_i$. For a given dimensionality $k$, $\bm{x}$ is drawn from the following distribution:
\begin{align*}
    \bm{x} &\sim \mathcal{N}(\bm{\mu_s}, \bm{\Sigma}) \\
    \bm{\mu}_s &= \mu_\textrm{target}\,\textbf{e}_s \\
    \bm{\Sigma} &= \bm{I}_k
\end{align*}
where $\textbf{e}_s$ is a $k$-dimensional basis vector with 1 on the $s$th coordinate and 0s elsewhere, $\mu_\textrm{target}$ is the observers' sensitivity to the target dimension, and $\bm{I}_k$ indicates the $k\times k$ identity matrix. Here we will focus on the case where $\mu_\textrm{target} = 1$, but similar results obtain for other values of $\mu_\textrm{target}$.

\begin{figure}[htbp]
    \centering
    \includegraphics[width=\linewidth]{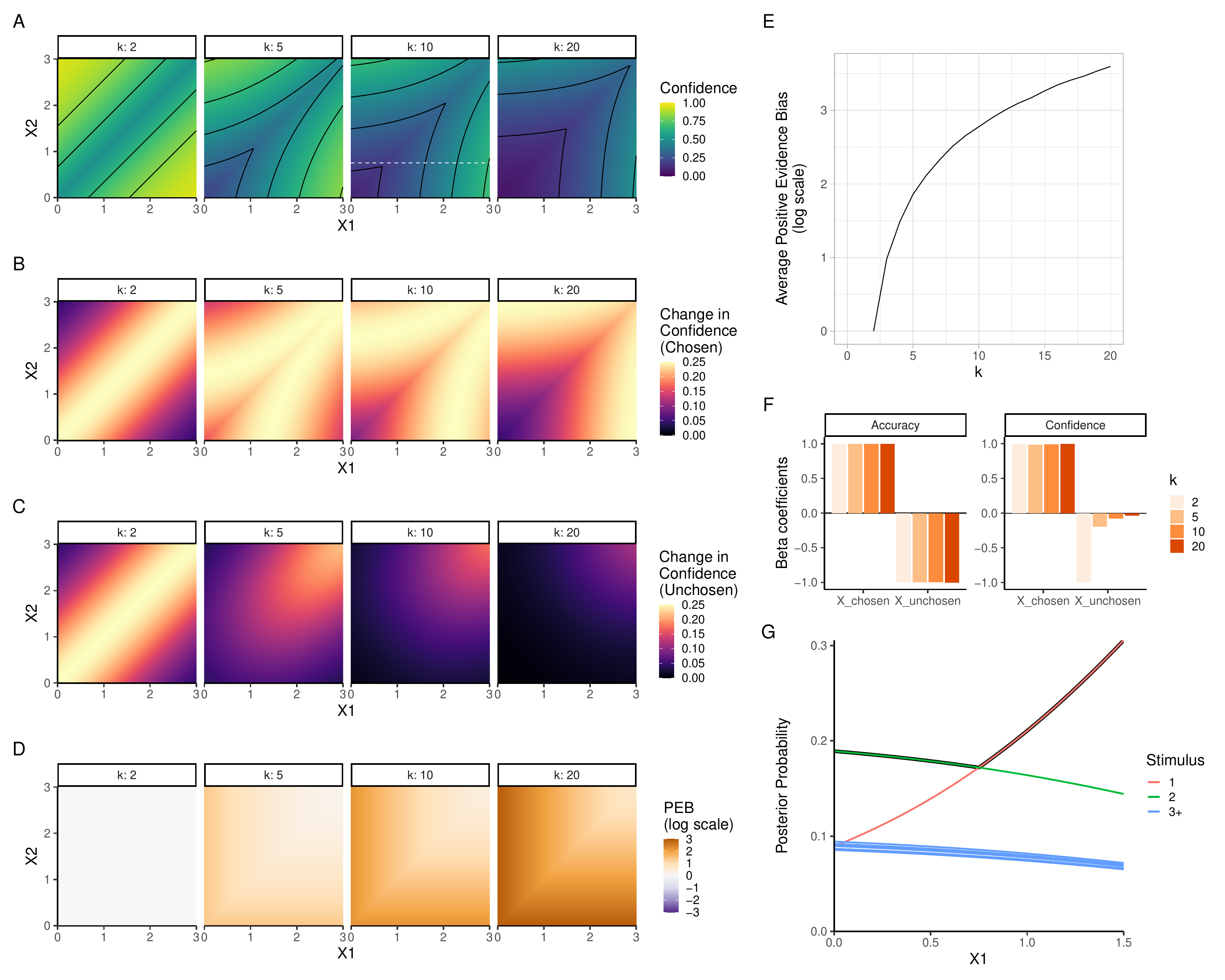}
    \caption{A-E) Impact of computing confidence in higher-dimensional SDT spaces for a 2D decision between $s_1$ and $s_2$ on (A) confidence, (B) confidence's sensitivity to evidence for the chosen alternative, (C) confidence's sensitivity to evidence for the unchosen alternative, and (D-E) the positive evidence bias (PEB), here computed as the log ratio of the two previous quantities. Confidence is increasingly detection-like (cf. Figure \ref{fig1}) as dimensionality increases, exhibiting increased sensitivity to evidence for the chosen alternative relative to the unchosen alternative. F) Regression coefficients relating chosen and unchosen evidence samples to accuracy (left panel) and confidence (right panel). G) Posterior probabilities of each stimulus class (colored lines) and confidence (black line) along the slice indicated by the dashed white line in panel (A) for $k=10$.}
    \label{fig2}
\end{figure}

As before, we assume the observer infers the stimulus identity by computing the posterior over all possible stimuli, $P(s_1 \vert \bm{x}) \ldots P(s_k \vert \bm{x})$. Then, the observer chooses the available action which maximizes this probability ($a = \argmax_{i\in \{1,2\}} P(s_i \vert \bm{x})$) and rates confidence as the posterior probability of the chosen stimulus ($c = P(s_a \vert \bm{x})$). The posterior probabilities can be computed using Bayes' rule:
\begin{equation}
\label{eq1}
c = P(s_a \vert \bm{x}) 
= \frac{P(s_a)p(\bm{x}\vert s_a)}{P(s_1)p(\bm{x}\vert s_1) + P(s_2)p(\bm{x}\vert s_2) + \ldots + P(s_k)p(\bm{x}\vert s_k)}
\end{equation}
Assuming a flat prior across stimuli (i.e., $P(s_1) = \ldots = P(s_k)$) we obtain:
\begin{equation}
\label{eq2}
c = P(s_a \vert \bm{x}) 
= \frac{p(\bm{x}\vert s_a)}{p(\bm{x}\vert s_1) + p(\bm{x}\vert s_2) + \ldots + p(\bm{x}\vert s_k)}
\end{equation}
As we show in Appendix \ref{appendix:confidence}, this equation can be solved to show that confidence can be straightforwardly computed as the softmax of the evidence with inverse temperature equal to the observer's sensitivity ($\beta = \mu_\textrm{target}$):
\begin{equation}
    \label{eq:softmax}
    \begin{split}
        c &= P(s_a\vert \bm{x}) \\
        &= \sigma_{\mu_\textrm{target}}(\bm{x})_a \\
        &= \frac{1}{e^{\mu_\textrm{target}(x_1-x_a)} + e^{\mu_\textrm{target}(x_2-x_a)} + \ldots + e^{\mu_\textrm{target}(x_k-x_a)}}
    \end{split}
\end{equation}
This equation emphasizes that, as $k$ increases, confidence is subject to increasing normalisation (for a similar derivation for detection tasks, see Appendix \ref{appendix:detection}). To plot out example surfaces in Figure \ref{fig2}, we set the unchosen evidence values to 0, but similar results are obtained when this is allowed to vary. All simulation code is available via the Open Science Foundation at \href{https://osf.io/uxphg}{https://osf.io/uxphg}.

\subsection{Neural network models}
To ask whether similar effects of dimensionality on confidence formation would obtain in more realistic stimulus sets, we trained the convolutional neural networks introduced by \citeauthor{Webb2023} (\citeyear{Webb2023}) to classify MNIST digits and report confidence (\href{https://github.com/taylorwwebb/performance_optimized_NN_confidence}{https://github.com/taylorwwebb/performance\_optimized\_NN\_confidence}).

The model architecture was the same as in \citet{Webb2023}, and involved three main components. A deep neural network encoder $f$ takes an input image $x$ of class $y$. The output of this encoder is then passed to two heads, $g_{class}$ and $g_{conf}$. The classification head $g_{class}$ outputs a predicted class for the image, while the confidence head $g_{conf}$ predicts $p(\hat{y}=y)$, the probability that the classification response is correct. The encoder $f$ consisted of three convolutional layers followed by three fully-connected (FC) layers. Each convolutional layer had 32 channels, a kernel size of 3, a stride of 2, batch normalisation, and leaky ReLU nonlinearities with a negative slope of 0.01. The first two FC layers had 256 and 128 units, respectively and included batch normalisation, and leaky ReLU nonlinearities with a negative slope of 0.01. The encoder output was produced by a final FC layer with 100 units and no nonlinearity or normalisation.

To manipulate dimensionality, we modified the number of distinct digits used as stimuli during network training. Dimensionality conditions ranging from $k=2$ to $k=10$ were created by randomly drawing subsets of $k$ digits without replacement from the set of integers between 0 and 9. Within each subset, we created high and low positive evidence exemplars by manipulating the signal-to-noise ratio of the stimuli, following the procedure in \citep{Webb2023}. Signal strength was varied by adjusting digit brightness from 0.1 to 1.0 (where black = 0 and white = 1), with noise superimposed as Gaussian-distributed brightness levels at each pixel in the image (both digit and background) with a standard deviation ranging between 0.0 and 2.0. Pixel values were demeaned before the noise was added, and the final values were truncated between $-1$ (black) and 1 (white). Two pairs of signal and noise levels were then established to create high and low positive evidence conditions. Noise level values were pre-established (low noise=1, high noise=2), and contrast levels (between 0 and 1) were adjusted to find High Positive Evidence (HPE) and Low Positive Evidence (LPE) conditions that resulted in equal discrimination accuracies of 55$\%$ correct. Each neural network was evaluated on the digit classes it was trained on, using held-out base images, under both HPE and LPE conditions.

\section{Results}
\subsection{Discrimination confidence criteria in higher dimensional spaces}

As shown in Figure \ref{fig2}A, as the dimensionality of the evidence space increases, confidence surfaces for a two-choice discrimination decision become more detection-like and more consistent with a PEB. This effect is due to the presence of additional ``distractor'' dimensions influencing confidence formation---if these dimensions could be ignored (more formally, if normalisation of confidence could be restricted to the 2D space), then we would continue to observe discrimination-like criteria even at higher dimensionalities.

We can quantify this by computing the absolute change in confidence with respect to the evidence for the chosen stimulus category $a$ and an unchosen stimulus category $i \ne a$, and taking the log ratio of the two partial derivatives. When this value is equal to 0, confidence is equally sensitive to changes in the chosen evidence $x_a$ and the unchosen evidence $x_i$. When this value is greater than 0, confidence is more sensitive to the chosen evidence $x_a$ relative to the unchosen evidence $x_i$, which is the hallmark of the positive evidence bias. In Appendix \ref{appendix:peb}, we show that the positive evidence bias takes the following form:
\begin{equation}
    \label{eq:peb}
    \textrm{PEB}_i = \textrm{log}\!\left\lvert \frac{\frac{\partial}{\partial x_a} P(s_a \vert \bm{x})}{\frac{\partial}{\partial x_i} P(s_a \vert \bm{x})} \right\rvert = \textrm{log}\!\left(1 + \sum_{j \notin \{a,i\}} e^{\mu_\textrm{target}(x_j-x_i)}\right)
\end{equation}
The strength of the PEB determined by Equation \ref{eq:peb} is plotted in Figure \ref{fig2}D. Since there is only one unchosen alternative for $k=2$, there are no terms in the sum, and so evidence for chosen and unchosen alternatives exhibit equal influences on confidence in this case ($\textrm{PEB}_i = \textrm{log}(1) = 0$). The sum gains an additional positive term for every increase in $k$, meaning that the PEB increases without bound as the number of distractor dimensions increases (Figure \ref{fig2}E). Importantly, this holds regardless of which choice the agent makes, regardless of which unchosen dimension we choose to investigate, and regardless of the actual evidence $\bm{x}$.

We next show that this effect of increasing dimensionality is sufficient to reproduce a classical PEB effect in simulation. Under a range of dimensionalities, we simulate evidence being drawn from $s_1$ or $s_2$, and record the model's choice and confidence. We then regressed the chosen and unchosen evidence against both choice and confidence using a logistic regression for choice and a linear regression on the logistic scale for confidence (given that confidence is bounded between 0 and 1; see Appendix \ref{appendix:logit}). As expected, choice was equally and oppositely affected by evidence for and against the chosen option across dimensionalities (Figure \ref{fig2}F). In contrast, confidence is driven to a greater extent by the chosen evidence, and this asymmetry increases as dimensionality increases.

\subsection{Explaining the influence of increasing dimensionality}

Why does this effect obtain? By taking a slice through the space at a constant, fixed value of $x_2$, and plotting the posterior for each dimension as a function of $x_1$, we can develop an intuition for this phenomenon (Figure \ref{fig2}G). When $x_1$ is low (and $s_1$ is unchosen), the other distractor dimensions strongly suppress its influence on confidence, leading to a weak influence of unchosen evidence. But when $x_1$ is high (when $s_1$ is chosen), it outcompetes the distractor dimensions, and confidence steeply rises with chosen evidence. This is due to the effect of an increasing number of distractor dimensions in the feature space entering into the normalisation term in Equation \ref{eq:softmax} as dimensionality ($k$) increases. In other words, a PEB naturally arises from a non-linearity in the function relating chosen evidence to confidence. This is solely due to the normalisation induced by the other feature dimensions in Equation \ref{eq:softmax}, rather than any bias or suboptimality in confidence formation.

\subsection{Revisiting variance asymmetries}

\citet{Miyoshi2020} proposed that a PEB may result from the evidence distributions in SDT having unequal variance along the target and non-target dimensions, leading to the adoption of curved, detection-like confidence criteria. However, their proposal does not explain why an inequality in variance may ensue in the first place. The current simulations in higher-dimensional spaces allow us to revisit this issue and obtain insight into one potential source of the variance inequality. 

\begin{figure}[htbp]
    \centering
    \includegraphics[width=.85\linewidth]{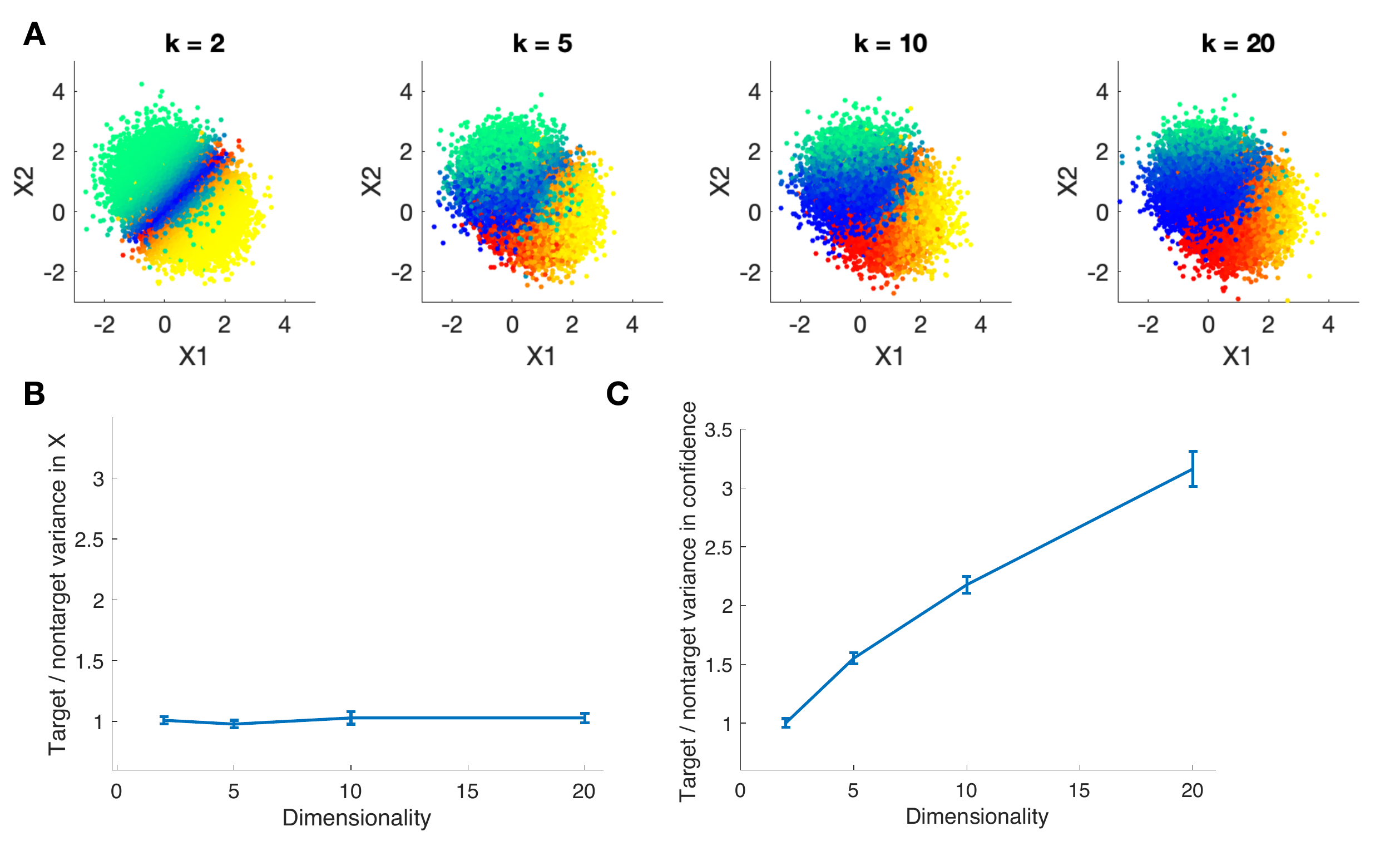}
    \caption{A) Samples from two equal variance stimulus distributions ($s_1$ and $s_2$), colour-coded by the level of confidence attached to a decision arising from each evidence sample, for increasing dimensionality. For a decision of $s_1$, confidence increases from blue to green. For a decision of $s_2$, confidence increases from red to yellow. B, C) Target:non-target variance ratio computed in (B) evidence space and (C) confidence space.}
    \label{fig3}
\end{figure}

In Figure \ref{fig3}A, we simulate samples from two equal variance stimulus distributions ($s_1$ and $s_2$), colour-coded by the decision and level of confidence arising from each evidence sample. In standard 2D SDT, all of the variance in confidence is on the minor diagonal, as expected, and the variances along the target and non-target dimensions are similar. But as dimensionality increases, the variance in confidence becomes greater in the target dimension, and weaker in the non-target dimension (this is a direct consequence of the confidence criteria becoming more detection-like, and parallel to the evidence axes). Importantly, however, this increasing variance in target (vs. non-target) confidence is obtained despite the \textit{evidence} distributions remaining equal in variance (the two clouds of points remain similar in shape and size across dimensionalities). This can be appreciated by computing the ratio of the variances along the major and minor axes as a function of dimensionality, computed either from the evidence samples (Figure \ref{fig3}B) or the confidence samples (Figure \ref{fig3}C). Only for the latter do we see the variance inequality increasing with dimensionality. In other words, the asymmetry in variance here is in the space of confidence, not in the space of evidence. We will revisit this observation in the Discussion, as this observation leads to an intriguing empirical prediction that it should be possible to ``flip'' a PEB (in confidence space) simply by changing the target dimension.

\subsection{Beyond the two-choice case}

\begin{figure}[htbp]
    \centering
    \includegraphics[width=\linewidth]{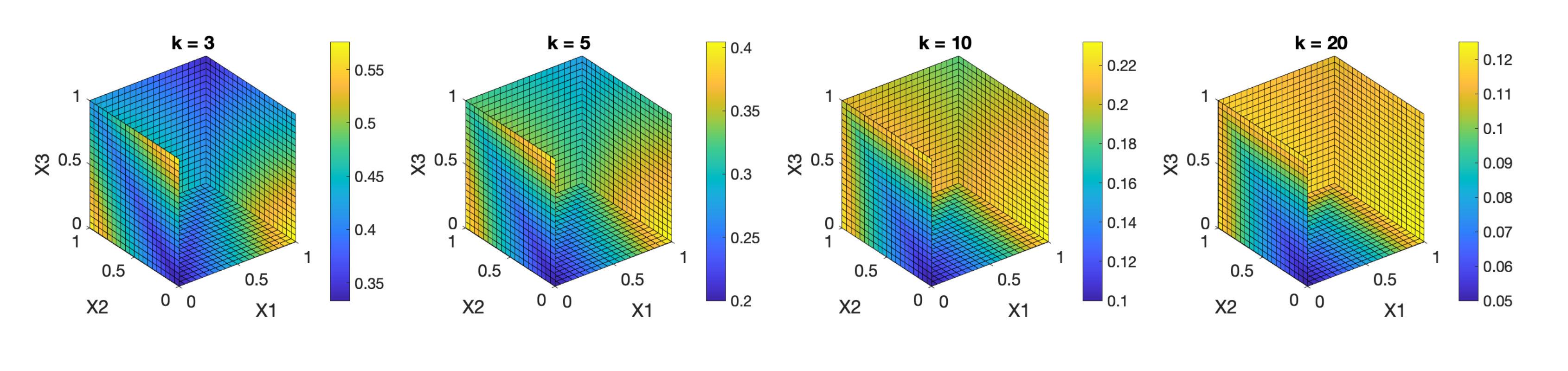}
    \caption{Impact of computing confidence in higher-dimensional SDT spaces on confidence surfaces for a 3-way decision between $s_1$, $s_2$ and $s_3$. Conventions as in Figure \ref{fig2}.}
    \label{fig4}
\end{figure}

So far we have considered a two-choice task faced by an observer using a higher-dimensional stimulus space to compute their (discrimination) confidence. But there is nothing special about the two-choice case. While it is harder to visualise, we can also demonstrate a similar effect in a 3-choice (or N-choice) task \citep{Li2020}, where dimensionality $k \geq N$. Figure \ref{fig4} shows the confidence surfaces for a 3-choice discrimination decision with increasing dimensionality of the evidence space. As for the 2-choice case, when $k=N$, the confidence criteria are aligned along the diagonal planes that separate one stimulus distribution from another. But as $k > N$, the confidence criteria become increasingly parallel to the sides of the cube---becoming increasingly detection-like.

\subsection{Effect of dimensionality on confidence formation in neural networks}

\begin{figure}[htbp]
    \centering
    \includegraphics[width=.66\linewidth]{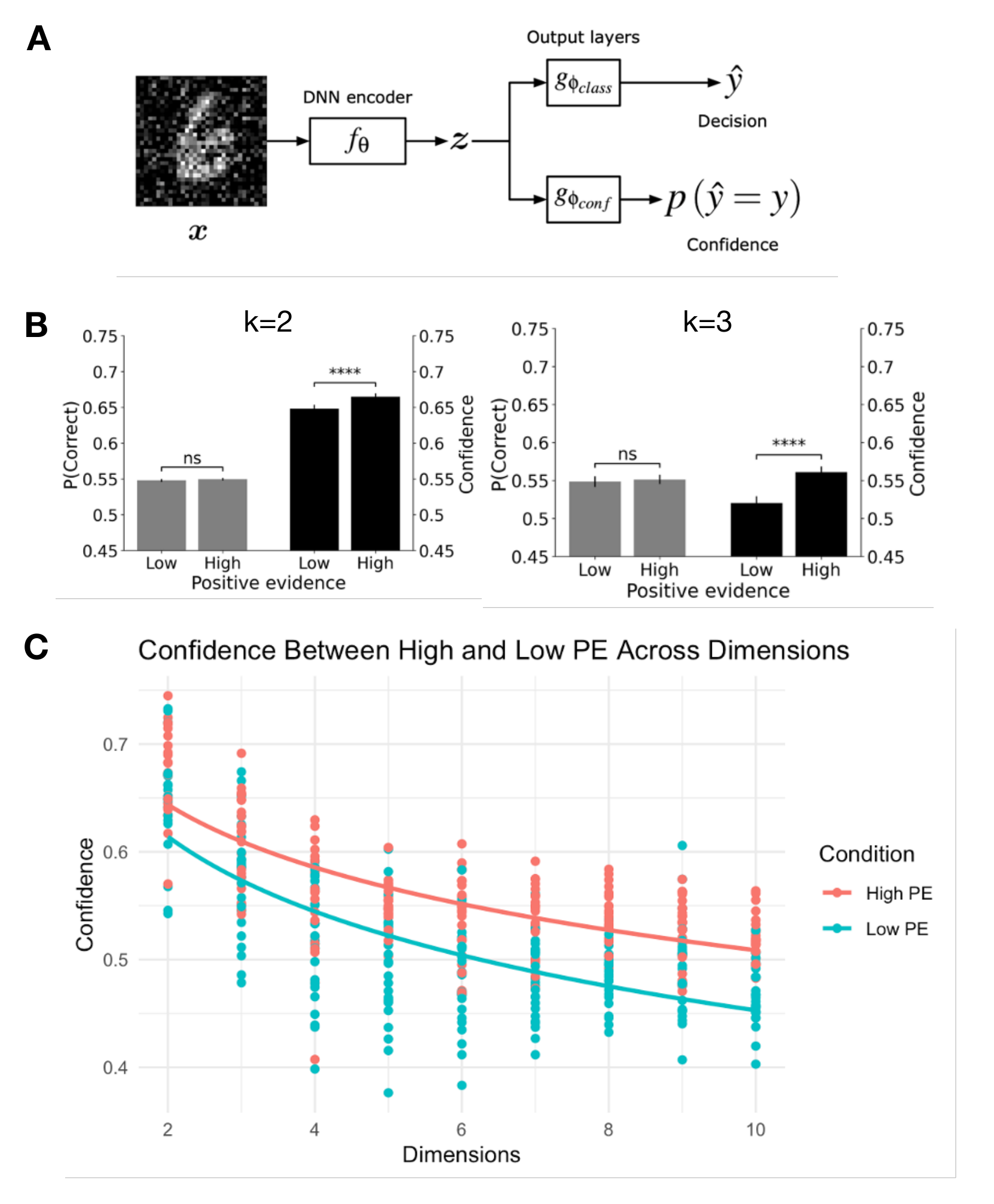}
    \caption{A) Model architecture \citep[modified from][]{Webb2023}. An image $x$, belonging to class $y$, was passed through a deep neural network (DNN) encoder $f$, followed by two output heads: $g_{class}$ generated a decision  classifying the image, and $g_{conf}$ generated a confidence score by predicting $p(\hat{y} = y)$, the probability that the decision was correct. B, C) Impact of varying the dimensionality of the training set on a positive evidence bias in confidence obtained for MNIST digit classification. The fitted lines are derived from a logarithmic model in which the impact of dimensionality on PEB saturates at higher dimensionalities.}
    \label{fig5}
\end{figure}

To ask whether similar effects of dimensionality on confidence formation would obtain in more realistic stimulus sets, we trained convolutional neural networks to classify MNIST digits and report confidence \citep[][Figure \ref{fig5}]{Webb2023}. Following \citet{Webb2023}, we created sets of HPE and LPE stimuli by simultaneously varying the signal and noise within each digit image (Figure \ref{fig5}). We varied the dimensionality of the stimulus space by varying how many (randomly subsampled) digits from 0-9 would be included in the network's training set (such that dimensionality $k$ ranged from 2 to 10).

As expected, when comparing high vs. low PE conditions, the network was more confident for high compared to low positive evidence stimuli, despite discrimination performance being clamped at 55$\%$ correct in both cases ($t(217) = -18.6, p <.001$). Strikingly, however, this effect of PE on confidence varied as a function of dimensionality (Figure \ref{fig5}B).

To quantify this effect, we built a linear mixed-effects model to ask how dimensionality affected the PEB. Confidence level was predicted by dimensionality, positive evidence (high vs. low) and their interaction, with random effects specified on the intercept for each individual neural network simulation. Increasing dimensionality tended to decrease confidence ($t(282.65)=-13.5, p<.0001$) whereas switching from low to high positive evidence increased confidence ($t(216)=13.53, p<.0001$), as expected. Notably, there was a significant interaction between dimensionality and PE ($t(216)=2.85, p=.005$), such that as dimensionality increased, a PEB also became more pronounced (Figure \ref{fig5}). Model comparison indicated that including the interaction term significantly improved the model fit over a nested model that did not contain the interaction term ($\chi^2(1)=8.04, p=0.0046$). Additional experiments revealed that this interaction remained significant across different training accuracies and was primarily sensitive to the dimensionality of the network's heads (rather than the encoder; see Appendix \ref{appendix:cnn}).

We also considered that the influence of dimensionality on the PEB may be better accommodated by a logarithmic, rather than linear relationship, due to the logarithmic form of Equation \ref{eq:peb} depicted in Figure \ref{fig2}E. This was indeed the case, with a logarithmic model providing a better fit as quantified with BIC scores (logarithmic, BIC = -6106.5; linear, BIC = -6092.5). The fits of the logarithmic model are included in Figure \ref{fig5}C.

\subsection{Dimensionality and metacognitive sensitivity}
\begin{figure}[htbp]
    \centering
    \includegraphics[width=\linewidth]{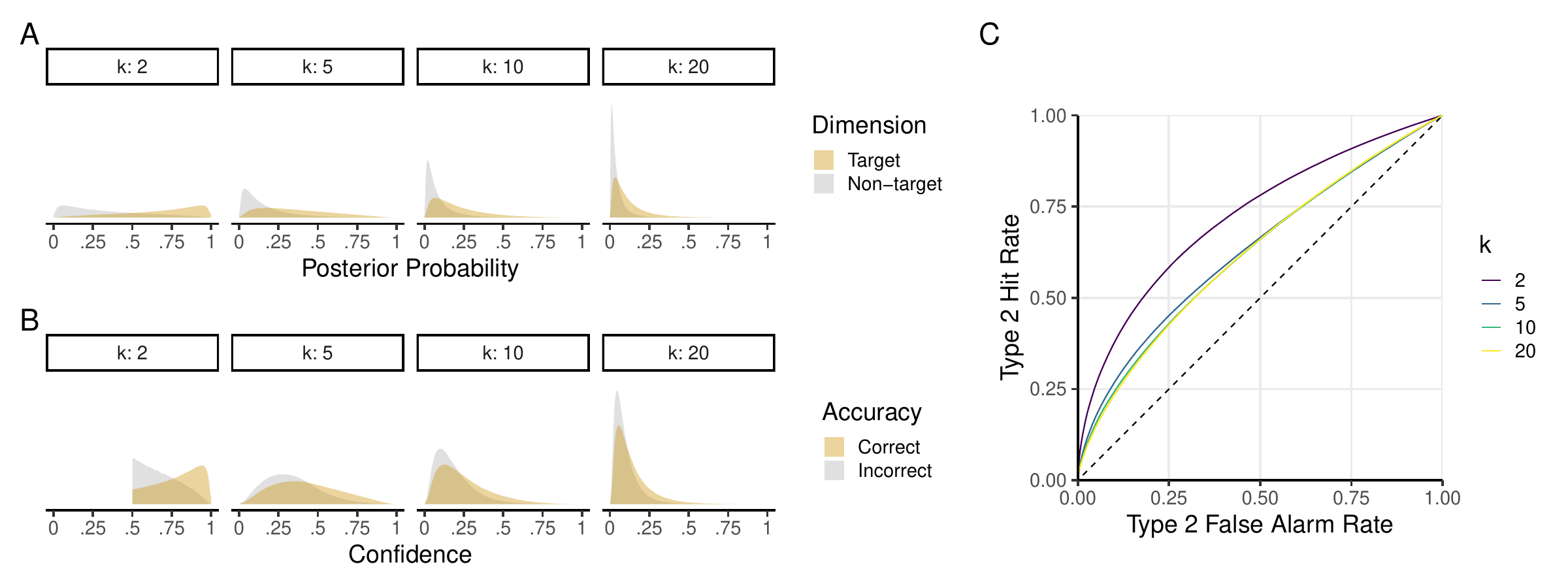}
    \caption{Influence of distractor dimensions on (A) the posterior probability of target and non-target stimuli, (B) confidence for correct and incorrect decisions, and (C) the type 2 ROC. For all values of $k$, posterior probabilities tend to be higher for target relative to non-target stimuli, confidence tends to be higher after a correct decision, and meta-d' is above 0. However, as $k$ increases, differences between target and non-target stimuli become less pronounced, resulting in more similar distributions of confidence for correct and incorrect decisions, and reduced metacognitive sensitivity.}
    \label{fig:metacognitive_sensitivity}
\end{figure}

We have shown that if participants consider a richer hypothesis space than assumed by the experimenter, the functional form of the optimal confidence criteria shifts, resulting in a positive evidence bias. We next asked whether dimensionality also affected metacognitive sensitivity---the extent to which confidence discriminates between correct and incorrect responses \citep{fleming2014}. In Figure \ref{fig:metacognitive_sensitivity}A-B, we plot distributions of posterior probabilities and confidence from simulated signal detection agents varying in their number of dimensions. As can be seen in Figure \ref{fig:metacognitive_sensitivity}A, although posterior probabilities always tend to be larger for target relative to non-target stimuli, these differences become smaller as $k$ increases. Likewise, in Figure \ref{fig:metacognitive_sensitivity}B, distributions of confidence for correct and incorrect decisions become more similar with increasing $k$. Figure \ref{fig:metacognitive_sensitivity}C plots the corresponding Type 2 ROCs for each value of $k$, which are defined by the Type 2 hit rate (i.e., the proportion of confidence ratings greater than $c$ following a correct decision) and the Type 2 false alarm rate (i.e., the proportion of confidence ratings greater than $c$ following an incorrect decision) across a range of confidence criteria $c$. Figure \ref{fig:metacognitive_sensitivity}C shows that the above pattern results in reduced metacognitive sensitivity, meaning that it is increasingly difficult to discriminate between correct and incorrect decisions in high-dimensional hypothesis spaces. Since dimensionality reduces metacognitive sensitivity but has no effect on Type 1 decision accuracy (see Appendix \ref{appendix:accuracy}), it follows that dimensionality also reduces metacognitive efficiency (the ratio of metacognitive sensitivity to Type 1 sensitivity) in a similar manner.

\section{Discussion}

Here we show that extending SDT into higher dimensional evidence spaces naturally leads to detection-like confidence surfaces when computing confidence in lower-dimensional (e.g., 2-choice) decisions. Our results extend the observations of \citet{Webb2023}, obtained as emergent properties of training neural networks on high-dimensional stimulus sets, to the framework of classical SDT. An advantage of this approach is that we are able to precisely isolate a source of the detection-like confidence effects in higher-dimensional spaces in the normalisation induced by other (distractor) stimulus dimensions on confidence.

Previous work has also sought to link the emergence of a positive evidence bias to the degree of normalisation in the computation of confidence. \citet{Maniscalco2021} proposed that the degree to which units within a neural network exhibit mutual inhibition predicts whether they will show discrimination- or detection-like confidence profiles. Less inhibition tuning led to detection-like confidence profiles, and accounted for empirical features of a positive evidence bias. This occurs because lower tuned inhibition leads to confidence being driven more by the absolute evidence than the difference in evidence between response options. In our SDT model, in contrast, detection-like confidence profiles are associated with \textit{greater} normalisation of confidence within a higher-dimensional feature space. These two perspectives can be reconciled by recognising that in \citet{Maniscalco2021}, the degree of inhibition is an intrinsic property of the network's units, rather than being a consequence of a confidence computation. It is possible, then, that a detection-like ``neuron'' that shows lower tuned inhibition within a neural network is in fact exactly what would be required to approximate the more aggressive algorithmic normalisation needed for computing (discrimination) confidence in higher-dimensional feature spaces.

Our model suggests that one source of unequal variance in signal detection models may be in the confidence computation itself, rather than (or in addition to) in the representation of evidence. We show in Appendix \ref{appendix:flip} that this insight is also consistent with recent findings that the positive evidence bias (and associated variance asymmetry) can be ``flipped'' just by changing the target dimension, with a positive evidence bias becoming a negative evidence bias if subjects are asked to report the \textit{lower} value item in a pair \citep{Sepulveda2020}. Such a flip would be more difficult to accommodate if a positive evidence bias was driven by lower level asymmetries in the representation of sensory evidence and/or a function of (signed) input magnitude \citep{Rausch2018}. Similar asymmetrical distributions have been observed in the latent space of neural network representations of the stimulus input \citep[obtained by using a variational autoencoder to extract a low-dimensional latent space underlying the high-dimensional stimulus inputs used in their experiments;][]{Webb2023}. These models capture unequal variance as a consequence of features of the training distribution, allowing them to explain how the PEB might arise in two-dimensional, stationary tasks. However, they presumably would not be able to accommodate an instantaneous flip in the PEB without retraining the model \citep{Sepulveda2020}. Here, as a complementary mechanism, we suggest that the rapid and flexible influence of changing the goal or hypothesis space being entertained by the subject is a key driver of the strength and sign of the PEB.

More generally, our results suggest, together with \citet{Webb2023}, that metacognitive biases may emerge due to subjects adopting a richer hypothesis space than assumed by the experimenter. This proposal aligns with other work showing how a mismatch between the experimenter's and subjects' task state-representations explains seemingly maladaptive behaviour in decision making \citep{Molano-Mazon2023} and metacognition \citep{Moore2023, Khalvati2021}. This hypothesis may be testable in future experiments that vary subjects' expectations about the task-relevance of different stimuli. For instance, it would be of interest to know whether a positive evidence bias becomes stronger when a subject is induced to expect a richer stimulus space, even if the range of stimuli in play remain constant. These and other hypotheses that emerge from extending SDT into higher dimensions make for a rich source of hypotheses for future experiments.

A broader implication of our results is that, in higher dimensional spaces, confidence criteria for detection and identification become more similar \citep{Mazor2023, Maniscalco2016}. In line with this idea, \citet{Mazor2023} showed that subjects incorporated detection-relevant evidence into their discrimination confidence, and discrimination-relevant evidence into their detection confidence. It is interesting to speculate that strict dissociations between detection and discrimination performance may be artefacts of the adoption of narrower stimulus spaces. Given that such dissociations have been foundational in establishing an empirical basis for unconscious perception \citep[e.g., in blindsight;][]{Weiskrantz1974, Azzopardi1997, Phillips2021, Peters2015, Peters2017}, establishing whether classical findings in consciousness science generalise to more naturalistic stimulus spaces represents an important direction for future research.

\section{Acknowledgments}
This work was supported by a UKRI/EPSRC Programme Grant [EP/V000748/1]. SMF is a CIFAR Fellow in the Brain, Mind \& Consciousness Program, and supported by UK Research and Innovation (UKRI) under the UK government's Horizon Europe funding guarantee (selected as ERC Consolidator, grant number 101043666). Parts of this work were previously presented at the Cognitive Computational Neuroscience Meeting, Oxford, August 2023.

\bibliography{LuczakFleming_highdimPE_PsychRev.bib}

\appendix
\section{Confidence can be computed using the softmax}\label{appendix:confidence}

As described in the Introduction, we can use Bayes' rule to obtain a posterior distribution over stimulus categories given the sensory evidence $\bm{x}$. Here we show that, assuming flat priors over stimulus categories, this reduces to the softmax of the evidence. We start by rewriting Equation \ref{eq2}:
\begin{equation}
\label{eq:confidence_rewrite}
c = P(s_a \vert \bm{x}) 
= \frac{p(\bm{x}\vert s_a)}{\sum_{i=1}^k p(\bm{x}\vert s_i)} = 
\frac{1}{\sum_{i=1}^k \frac{p(\bm{x}\vert s_i)}{p(\bm{x}\vert s_a)}}
\end{equation}
For any stimulus $s$, $p(\bm{x} \vert s)$ is the likelihood of the joint normal distribution over $\bm{x}$ with mean $\bm{\mu}_s = \mu_\textrm{target}\,\textbf{e}_s$ and covariance $\bm{\Sigma} = \bm{I}_k$, where $\bm{\mu}_s$ is a $k$-dimensional vector with $\mu_\textrm{target}$ on the $s$th coordinate and 0s elsewhere, $\mu_\textrm{target}$ is the observers' sensitivity to the target dimension, and $\bm{I}_k$ indicates the $k\times k$ identity matrix. The density of the multivariate normal distribution with covariance $\bm{\Sigma} = \bm{I}_k$ is defined as:
\begin{align*}
    p(\bm{x}\vert\bm{\mu}_s,\bm{\Sigma}) &= \frac{1}{\sqrt{(2\pi)^k\vert\bm{\Sigma}\vert}}e^{-\frac{1}{2}(\bm{x} - \bm{\mu}_s)^\top\bm{\Sigma}^{-1}(\bm{x} - \bm{\mu}_s)} \\
    &= \frac{1}{\sqrt{(2\pi)^k}}e^{-\frac{1}{2}\|\bm{x} - \bm{\mu}_s\|^2}
\end{align*}
We can plug this into Equation \ref{eq:confidence_rewrite} and simplify to derive confidence:
\begin{align*}
    c &= P(s_a \vert \bm{x}) \\
    &= \frac{1}{\sum_{i=1}^k \frac{p(\bm{x}\vert s_i)}{p(\bm{x}\vert s_a)}} \\
    &= \frac{1}{\sum_{i=1}^k \frac{e^{-\frac{1}{2}\|\bm{x} - \bm{\mu}_{s_i}\|^2}}{e^{-\frac{1}{2}\|\bm{x} - \bm{\mu}_{s_a}\|^2}}} \\
    &= \frac{1}{\sum_{i=1}^k e^{-\frac{1}{2}\left[\|\bm{x} - \bm{\mu}_{s_i}\|^2 - \|\bm{x} - \bm{\mu}_{s_a}\|^2\right]}} \\
    &= \frac{1}{\sum_{i=1}^k e^{-\frac{1}{2}\left[(x_i - \mu_\textrm{target})^2 + \sum_{j\ne i} x_j^2 - (x_a - \mu_\textrm{target})^2 - \sum_{j\ne a} x_j^2\right]}} \\
    &= \frac{1}{\sum_{i=1}^k e^{-\frac{1}{2}\left[
    (x_i - \mu_\textrm{target})^2 + x_a^2 + \sum_{j\notin\{a,i\}}x_j^2 - (x_a - \mu_\textrm{target})^2 - x_i^2 - \sum_{j\notin\{a,i\}}x_j^2 \right]}} \\
    &= \frac{1}{\sum_{i=1}^k e^{\mu_\textrm{target}(x_i - x_a)}} \\
    &= \frac{e^{\mu_\textrm{target} \, x_a}}{\sum_{i=1}^k e^{\mu_{\textrm{target}} \, x_i}} \\
    &= \sigma_{\mu_\textrm{target}}(\bm{x})_a
\end{align*}
As a result, the full posterior distribution over all possible stimuli $\bm{s} = \left[ s_1 \ldots s_k \right]$ can be efficiently calculated using the softmax with inverse temperature $\beta = \mu_\textrm{target}$:
\begin{equation}
    P(\bm{s}\vert \bm{x}) =
    \begin{bmatrix}
       P(s_1\vert \bm{x}) \\
       \vdots \\
       P(s_k\vert \bm{x})
    \end{bmatrix} = \sigma_{\mu_\textrm{target}}(\bm{x})
\end{equation}

\section{The positive evidence bias grows with dimensionality}\label{appendix:peb}
The positive evidence bias is defined as an increased sensitivity in confidence to the evidence for the chosen alternative, relative to evidence for unchosen alternatives. To quantify the positive evidence bias, then, we need to take the partial derivative of confidence with respect to each of the evidence dimensions $x_i$:
\begin{align*}
    \frac{\partial}{\partial x_i} P(s_a \vert \bm{x}) &= \frac{\partial}{\partial x_i} \sigma_{\mu_\textrm{target}}(\bm{x})_a \\
    &= \frac{\partial}{\partial x_i} \frac{e^{\mu_\textrm{target}\,x_a}}{\sum_{j} e^{\mu_\textrm{target}\,x_j}} \\
    &= \frac{\frac{\partial}{\partial x_i}\left[e^{\mu_\textrm{target}\,x_a}\right]\sum_{j} e^{\mu_\textrm{target}\,x_j} - e^{\mu_\textrm{target}\,x_a}\frac{\partial}{\partial x_i}\left[\sum_{j} e^{\mu_\textrm{target}\,x_j}\right]}{(\sum_{j} e^{\mu_\textrm{target}\,x_j})^2}
\end{align*}
For $i = a$, this reduces to:
\begin{align*}
    \frac{\partial}{\partial x_i} P(s_a \vert \bm{x}) &= \frac{\mu_\textrm{target} \, e^{\mu_\textrm{target}\,x_a} \, \sum_{j} e^{\mu_\textrm{target}\,x_j} - \mu_\textrm{target} \, e^{\mu_\textrm{target}\,x_a} \, e^{\mu_\textrm{target}\,x_a}}{(\sum_{j} e^{\mu_\textrm{target}\,x_j})^2} \\
    &= \mu_\textrm{target} \, \frac{e^{\mu_\textrm{target}\,x_a}}{\sum_{j} e^{\mu_\textrm{target}\,x_j}} \, \frac{\sum_{j} e^{\mu_\textrm{target}\,x_j} - e^{\mu_\textrm{target}\,x_a}}{\sum_{j} e^{\mu_\textrm{target}\,x_j}} \\
    &= \mu_\textrm{target} \, \sigma_{\mu_\textrm{target}}(\bm{x})_a \, (1 - \sigma_{\mu_\textrm{target}}(\bm{x})_a) \\
    &= \mu_\textrm{target} \, P(s_a\vert\bm{x}) \, (1 - P(s_a\vert\bm{x}))
\end{align*}
For $i \ne a$, it reduces to:
\begin{align*}
    \frac{\partial}{\partial x_i} P(s_a \vert \bm{x}) &= \frac{-\mu_\textrm{target} \, e^{\mu_\textrm{target}\,x_a} \, e^{\mu_\textrm{target}\,x_i}}{(\sum_{j} e^{\mu_\textrm{target}\,x_j})^2} \\
    &= -\mu_\textrm{target}\,\sigma_{\mu_\textrm{target}}(x)_a\,\sigma_{\mu_\textrm{target}}(x)_i \\
    &= -\mu_\textrm{target}\, P(s_a\vert\bm{x}) \, P(s_i\vert\bm{x})
\end{align*}
Together, these define the Jacobian matrix $\bm{J}_{P(\bm{s} \vert \bm{x})} = \mu_\textrm{target}\left[\textrm{diag}(\sigma_{\mu_\textrm{target}}(\bm{x})) - \sigma_{\mu_\textrm{target}}(\bm{x})\sigma_{\mu_\textrm{target}}(\bm{x})^\top\right]$. Since the PEB occurs when the absolute change in confidence is greater with respect to $x_a$ than $x_i$ for $i \ne a$, we can determine this using the log ratio of the two partial derivatives:
\begin{align*}
    \textrm{PEB}_i &= \textrm{log}\!\left\lvert\frac{\frac{\partial}{\partial x_a} P(s_a \vert \bm{x})}{\frac{\partial}{\partial x_i} P(s_a \vert \bm{x})} \right\rvert \\
    &= \textrm{log}\!\left(\frac{\mu_\textrm{target} \, \sigma_{\mu_\textrm{target}}(\bm{x})_a \, (1 - \sigma_{\mu_\textrm{target}}(\bm{x})_a)}{\mu_\textrm{target} \, \sigma_{\mu_\textrm{target}}(\bm{x})_a \, \sigma_{\mu_\textrm{target}}(\bm{x})_i}\right) \\
    &= \textrm{log}\!\left(\frac{1 - \sigma_{\mu_\textrm{target}}(\bm{x})_a}{\sigma_{\mu_\textrm{target}}(\bm{x})_i}\right) \\
    &= \textrm{log}\!\left(\frac{1 - \frac{e^{\mu_\textrm{target} \, x_a}}{\sum_{j} e^{\mu_\textrm{target} \, x_j}}}{\frac{e^{\mu_\textrm{target} \, x_i}}{\sum_{j} e^{\mu_\textrm{target} \, x_j}}}\right) \\
    &= \textrm{log}\!\left(\frac{\sum_{j} e^{\mu_\textrm{target} \, x_j} - e^{\mu_\textrm{target} \, x_a}}{e^{\mu_\textrm{target} \, x_i}}\right) \\
    &= \textrm{log}\!\left(\sum_{j} e^{\mu_\textrm{target}(x_j-x_i)} - e^{\mu_\textrm{target}(x_a-x_i)}\right) \\
    &= \textrm{log}\!\left(1 + \sum_{j \notin \{a, i\}} e^{\mu_\textrm{target}(x_j-x_i)}\right)
\end{align*}

When $\textrm{PEB}_i = 0$, evidence for the chosen and unchosen options exert the same influence on confidence. The positive evidence bias arises when $\textrm{PEB}_i > 0$, since in this case the evidence for the chosen option exerts a larger influence than the evidence for an unchosen option. $\textrm{PEB}_i = 0$ for $k=2$, since there are no unchosen alternatives other than $i$, and so there are no terms in the sum. For $k > 2$, $\textrm{PEB}_i > 0$, since each additional term in the sum must be positive. If we take the limit as $k \rightarrow \infty$, we obtain:
\begin{align*}
    \lim_{k\to\infty} \textrm{PEB}_i
    = \lim_{k\to\infty} \textrm{log}\!\left(1 + \sum_{j \notin \{a, i\}} e^{\mu_\textrm{target}(x_j-x_i)}\right) = \infty
\end{align*}
So, for sufficiently large $k$, the influence of evidence for the chosen option on confidence will dwarf the influence of evidence for unchosen options, leading to detection-like confidence criteria.

\section{Dimensionality does not affect decision accuracy}\label{appendix:accuracy}
On the basis of the $k$-dimensional signal $\bm{x}$, the signal detection agent chooses an action that maximizes the posterior probability: $a = \argmax_{i\in \{1,2\}} P(s_i \vert \bm{x})$.
Without loss of generality, assume the true stimulus is $s=s_1$ such that $x_1 \sim \mathcal{N}(\mu_\textrm{target}, 1)$ and $x_{2:k} \sim \mathcal{N}(0, 1)$. The agent makes a correct decision whenever
\begin{align*}
    P(s_1 \vert \bm{x}) &> P(s_2 \vert \bm{x}) \\
    \sigma_{\mu_\textrm{target}}(\bm{x})_1 &> \sigma_{\mu_\textrm{target}}(\bm{x})_2 \\
    \frac{e^{\mu_\textrm{target} \, x_1}}{\sum_{i=1}^k e^{\mu_{\textrm{target}} \, x_i}} &> \frac{e^{\mu_\textrm{target} \, x_2}}{\sum_{i=1}^k e^{\mu_{\textrm{target}} \, x_i}} \\
    0 &> x_2 - x_1 \\
\end{align*}

Since $x_1$ and $x_2$ are independently normally distributed, it follows that $x_2 - x_1 \sim \mathcal{N}(-\mu_\textrm{target}, \sqrt{2})$. Then the agent's accuracy is:
\begin{align*}
    P(x_1 > x_2) &= P(x_2-x_1 < 0) \\
    &= \Phi\left(\frac{\mu_\textrm{target}}{\sqrt{2}}\right)
\end{align*}
This formula makes clear that the agent's accuracy depends only on $\mu_\textrm{target}$ and not on dimensionality ($k$).

\section{Effect of dimensionality on the positive evidence bias on logit scale}\label{appendix:logit}
Since confidence is bounded between 0 and 1, it is sometimes easier to see the effects on confidence on the logistic scale (i.e., by analyzing $\textrm{logit}(c) = \textrm{log}\frac{c}{1-c}$). First, we can express confidence on this scale:
\begin{align*}
    \textrm{logit}(c) &= \textrm{log}\left(\frac{P(s_a \vert \bm{x})}{1-P(s_a \vert \bm{x})}\right) \\
    &= \textrm{log}\left(\frac{\frac{e^{\mu_\textrm{target}x_a}}{\sum_{i=1}^k e^{\mu_\textrm{target}x_i}}}{1-\frac{e^{\mu_\textrm{target}x_a}}{\sum_{i=1}^k e^{\mu_\textrm{target}x_i}}}\right) \\
    &= \textrm{log}\left(\frac{e^{\mu_\textrm{target}x_a}}{\sum_{i=1}^k e^{\mu_\textrm{target}x_i}-e^{\mu_\textrm{target}x_a}}\right) \\
    &= \mu_\textrm{target}x_a - \textrm{log}\left(\sum_{i \ne a} e^{\mu_\textrm{target}x_i}\right) \\
\end{align*}

As before, we want to see the effect of evidence for chosen and unchosen options on confidence. By taking the partial derivative of confidence with respect to $x_a$, we see that the effect of chosen evidence on confidence is just $\mu_\textrm{target}$:
\begin{align*}
    \frac{\partial}{\partial x_a} \textrm{logit}(c) &= \frac{\partial}{\partial x_a} \left[ \mu_\textrm{target}x_a - \textrm{log}\left(\sum_{i \ne a} e^{\mu_\textrm{target}x_i}\right) \right] \\
    &= \frac{\partial}{\partial x_a} \left[ \mu_\textrm{target}x_a\right] - \frac{\partial}{\partial x_a}\left[\textrm{log}\left(\sum_{i \ne a} e^{\mu_\textrm{target}x_i}\right) \right] \\
    &= \mu_\textrm{target}
\end{align*}

Through a similar process, we can see that the effect of evidence for an unchosen option $x_i$ on confidence is $-\mu_\textrm{target}$, scaled by the relative evidence for $x_i$ over the other unchosen options: 
\begin{align*}
    \frac{\partial}{\partial x_i} \textrm{logit}(c) &= \frac{\partial}{\partial x_i} \left[ \mu_\textrm{target}x_a - \textrm{log}\left(\sum_{j \ne a} e^{\mu_\textrm{target}x_j}\right) \right] \\
    &= \frac{\partial}{\partial x_i} \left[ \mu_\textrm{target}x_a\right] - \frac{\partial}{\partial x_i}\left[\textrm{log}\left(\sum_{j \ne a} e^{\mu_\textrm{target}x_j}\right) \right] \\
    &= -\mu_\textrm{target}\frac{e^{\mu_\textrm{target}x_i}}{\sum_{j\ne a} e^{\mu_\textrm{target}x_j}} \\
    &= -\mu_\textrm{target}\frac{\frac{e^{\mu_\textrm{target}x_i}}{\sum_{j} e^{\mu_\textrm{target}\,x_j}}}{\frac{\sum_{j\ne a} e^{\mu_\textrm{target}x_j}}{\sum_{j} e^{\mu_\textrm{target}\,x_j}}} \\
    &= -\mu_\textrm{target}\frac{\sigma(\bm{x})_i}{1 - \sigma(\bm{x})_a} \\
    &= -\mu_\textrm{target}\frac{P(s_i \vert \bm{x})}{1 - P(s_a \vert \bm{x})}
\end{align*}

From here we can see that the effect of chosen and unchosen evidence on $\textrm{logit}(c)$ are the same as for $c$, except divided by $P(s_a\vert\bm{x})(1 - P(s_a\vert\bm{x}))$. Thus we obtain the same positive evidence bias on the logit scale:
\begin{align*}
    \textrm{PEB}_i &= \textrm{log}\!\left\lvert\frac{\frac{\partial}{\partial x_a} \textrm{logit}(P(s_a \vert \bm{x}))}{\frac{\partial}{\partial x_i} \textrm{logit}(P(s_a \vert \bm{x}))} \right\rvert \\
    &= \textrm{log}\!\left\lvert\frac{\mu_\textrm{target}}{-\mu_\textrm{target}\frac{P(s_i \vert \bm{x})}{1 - P(s_a \vert \bm{x})}} \right\rvert \\
    &= \textrm{log}\!\left( \frac{\sum_{j\ne a} e^{\mu_\textrm{target}x_j}}{e^{\mu_\textrm{target}x_i}} \right) \\
    &= \textrm{log}\!\left( \sum_{j\ne a} e^{\mu_\textrm{target} (x_j - x_i)} \right) \\
    &= \textrm{log}\!\left( 1 + \sum_{j\notin \{a,i\}} e^{\mu_\textrm{target} (x_j - x_i)} \right) \\
\end{align*}

\section{Positive evidence bias flips with task demands}\label{appendix:flip}
Now assume that instead of choosing the stimulus with the strongest evidence, the observer is tasked with choosing the stimulus with the weakest evidence \citep{Sepulveda2020}. In this setting, confidence is be equal to the complement of the posterior probability of the chosen stimulus:
\begin{align*}
    a &= \argmin_{i\in \{1,2\}} P(s_i \vert \bm{x}) \\
    c &= 1 - P(s_a \vert \bm{x})
\end{align*}
We can determine the PEB in a similar manner as before:
\begin{align*}
    \textrm{PEB}_i &= \textrm{log}\!\left\lvert\frac{\frac{\partial}{\partial x_i} \left[1-P(s_a \vert \bm{x})\right]}{\frac{\partial}{\partial x_a} \left[1-P(s_a \vert \bm{x})\right]} \right\rvert \\
    &= \textrm{log}\!\left\lvert\frac{\frac{\partial}{\partial x_i} P(s_a \vert \bm{x})}{\frac{\partial}{\partial x_a} P(s_a \vert \bm{x})} \right\rvert \\
    &= \textrm{log}\!\left(\frac{1}{1 + \sum_{j \notin \{a, i\}} e^{\mu_\textrm{target}(x_j-x_i)}}\right) \\
    &= - \textrm{log}\!\left(1 + \sum_{j \notin \{a, i\}} e^{\mu_\textrm{target}(x_j-x_i)}\right)
\end{align*}
So, when asked to report the stimulus with the weakest evidence, it is normative for confidence to depend on the evidence for the unchosen options more than the chosen option.

\section{High-dimensional signal detection for detection tasks}\label{appendix:detection}
Just as we can consider increasing the size of the hypothesis space for discrimination tasks, we can do the same for detection tasks. In particular, we can consider a stimulus $s \in \{s_0, s_1, \ldots, s_k\}$ with presence $p \in \{p_0, p_1\}$, where $s_0$ represents an absent stimulus (when $p=p_0$) and $s_1\ldots s_k$ represent different possible stimuli when a stimulus is present (when $p=p_1$). In this setting, we assume the evidence $\bm{x}$ is sampled from the following distribution:
\begin{align*}
    \bm{x} &\sim \mathcal{N}(\bm{\mu_s}, \bm{\Sigma}) \\
    \bm{\mu}_s &= \begin{cases}
        \bm{0} & \textrm{if } p=p_0 \\
        \mu_\textrm{target}\,\textbf{e}_s & \textrm{otherwise}
    \end{cases} \\
    \bm{\Sigma} &= \bm{I}_k
\end{align*}
Then, the observer selects an action indicating the presence or absence of the stimulus and rates confidence according to the posterior probability that the decision is correct:
\begin{align*}
    a &= \begin{cases}
        0 & \text{if }P(p_0\vert\bm{x}) > P(p_1\vert\bm{x}) \\
        1 & \text{otherwise}
    \end{cases} \\
    c &= P(p_a \vert \bm{x})
\end{align*}

\begin{figure}
    \centering
    \includegraphics[width=\linewidth]{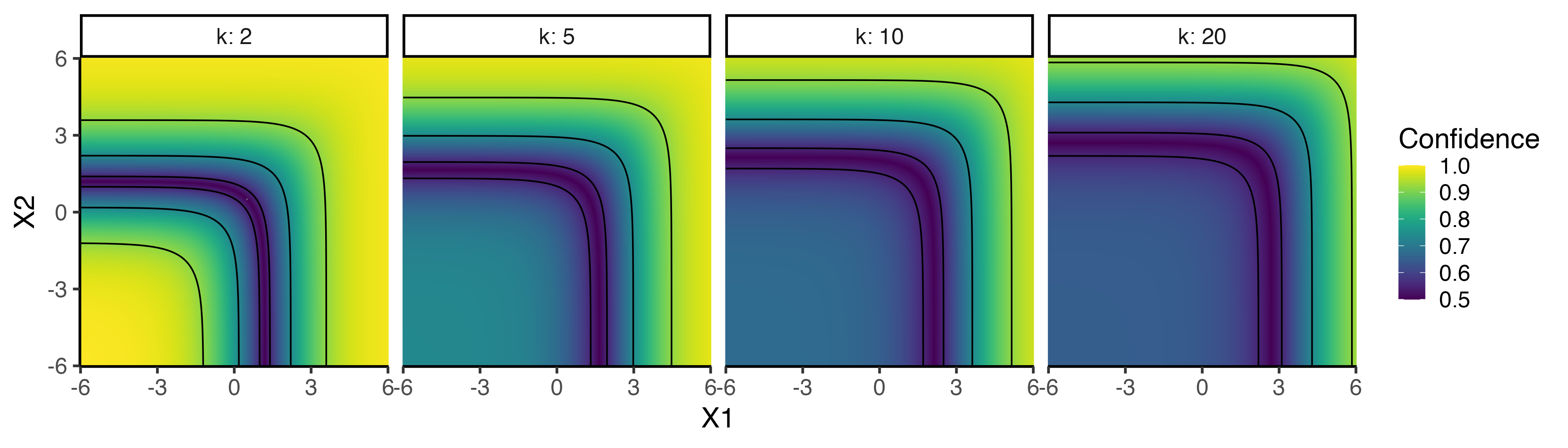}
    \caption{Impact on confidence of computing confidence in higher-dimensional SDT spaces for a detection task between $s = s_0$ (absent) and $s \in \{s_1 \ldots s_K\}$ (present).}
    \label{fig:detection}
\end{figure}

As before, the posterior probability can be computed using Bayes' rule. Without loss of generality (since $P(p_1\vert \bm{x}) = 1 - P(p_0\vert\bm{x})$), we can consider confidence that the stimulus is absent ($p=p_0$):
\begin{align*}
    P(p_0 \vert \bm{x})
    &= \frac{P(p_0)p(\bm{x}\vert p_0)}{P(p_0)p(\bm{x}\vert p_0) + P(p_1)p(\bm{x}\vert p_1)} \\
    &= \frac{P(p_0)p(\bm{x}\vert p_0)}{P(p_0)p(\bm{x}\vert p_0) + P(p_1)\Sigma_{i=1}^k p(\bm{x}\vert p_1,s_i)P(s_i\vert p_1)} \\
\end{align*}
Assuming equal prior probabilities of stimulus presence and absence ($P(p_0) = P(p_1) = \frac{1}{2}$) and equal probabilities of each stimulus given the presence of a stimulus ($P(s_1\vert p_1) = \ldots = P(s_k\vert p_1) = \frac{1}{k}$), we have: 
\begin{align*}
    P(p_0 \vert \bm{x})
    &= \frac{p(\bm{x}\vert p_0)}{p(\bm{x}\vert p_0) + \frac{1}{k}\sum_{i=1}^k p(\bm{x}\vert p_1,s_i)}\\
    &= \frac{p(\bm{x}\vert s_0)}{p(\bm{x}\vert s_0) + \frac{1}{k}\sum_{i=1}^k p(\bm{x}\vert s_i)} \\
    &= \frac{1}{1 + \frac{1}{k}\sum_{i=1}^k \frac{p(\bm{x}\vert s_i)}{p(\bm{x}\vert s_0)}} \\
\end{align*}
Plugging in the density of the bivariate normal distribution, we can derive confidence in the absence of a stimulus:
\begin{align*}
    P(p_0 \vert \bm{x}) 
    &= \frac{1}{1 + \frac{1}{k}\sum_{i=1}^k \frac{p(\bm{x}\vert s_i)}{p(\bm{x}\vert s_0)}} \\
    &= \frac{1}{1 + \frac{1}{k}\sum_{i=1}^k \frac{e^{-\frac{1}{2}\|\bm{x} - \bm{\mu}_{s_i}\|^2}}{e^{-\frac{1}{2}\|\bm{x} - \bm{\mu}_{s_0}\|^2}}} \\
    &= \frac{1}{1 + \frac{1}{k}\sum_{i=1}^k e^{-\frac{1}{2}\left[\|\bm{x} - \bm{\mu}_{s_i}\|^2 - \|\bm{x}\|^2\right]}} \\
    &= \frac{1}{1 + \frac{1}{k}\sum_{i=1}^k e^{-\frac{1}{2}\left[(x_i - \mu_\textrm{target})^2 + \sum_{j\ne i} x_j^2 - \sum_j x_j^2\right]}} \\
    &= \frac{1}{1 + \frac{1}{k}\sum_{i=1}^k e^{-\frac{1}{2}\left[x_i^2 - 2x_i\mu_\textrm{target} + \mu_\textrm{target}^2 - x_i^2\right]}} \\
    &= \frac{1}{1 + \frac{1}{k}\sum_{i=1}^k e^{x_i\mu_\textrm{target} -\frac{1}{2}\mu_\textrm{target}^2}} \\
    &= \frac{e^{\frac{1}{2}\mu_\textrm{target}^2}}{e^{\frac{1}{2}\mu_\textrm{target}^2} + \frac{1}{k}\sum_{i=1}^k e^{x_i\mu_\textrm{target}}} \\
\end{align*}
From this formula it is clear that the observer should report ``absent'' whenever $e^{\frac{1}{2}\mu_\textrm{target}^2} > \frac{1}{k}\sum_{i=1}^k e^{x_i\mu_\textrm{target}}$ and ``present'' otherwise. Through simulating this model, we found that that dimensionality had little effect on the qualitative shape of the confidence criteria (Figure \ref{fig:detection}). This is because, assuming evidence is close to zero, increasing $k$ does not change the average amount of evidence $\frac{1}{k}\sum_{i=1}^k e^{x_i\mu_\textrm{target}}$.

Dimensionality does produce two notable effects on confidence in a detection task, however. First, confidence in ``absent'' responses decreases with dimensionality. This effect occurs because, as more and more stimuli become possible, evidence against $s_1$ and $s_2$ is no longer sufficient to decide that $s=s_0$ (i.e., it could be that $s \in \{s_3 \ldots s_k\}$ despite the fact that each has low evidence). This is in contrast to the lack of change in confidence for ``present'' responses, since high values of $x_1$ or $x_2$ directly implicate the presence of $s_1$ or $s_2$, respectively.

Second, the curve at which confidence is lowest shifts along the diagonal towards the upper right  as dimensionality increases. This effect occurs because we have fixed evidence for distractor dimensions to zero. Specifically, confidence is at minimum ($c=.5$) when $e^{\frac{1}{2}\mu_\textrm{target}^2} \approx \frac{1}{k}\sum_{i=1}^k e^{x_i\mu_\textrm{target}}$. Assuming that $x_1=x_2=x$ and that $x_3 = \ldots = x_k = 0$, we can solve for the point along the diagonal with minimal confidence:
\begin{align*}
    e^{\frac{1}{2}\mu_\textrm{target}^2} &= \frac{1}{k}\sum_{i=1}^k e^{x_i\mu_\textrm{target}} \\
    1 &= \frac{1}{k}\sum_{i=1}^k e^{x_i\mu_\textrm{target} - \frac{1}{2}\mu_\textrm{target}^2} \\
    k &= 2e^{x\mu_\textrm{target} - \frac{1}{2}\mu_\textrm{target}^2} + (k-2)e^{- \frac{1}{2}\mu_\textrm{target}^2} \\
    x &= \frac{1}{\mu_\textrm{target}}\textrm{log}\left(\frac{1}{2}ke^{\frac{1}{2}\mu_\textrm{target}^2} - \frac{1}{2}k + 1\right) \\
\end{align*}
This expression makes clear that the point of minimal confidence is a monotonically increasing function of $k$. But this second effect is specific to our assumption that $x_3 = \ldots = x_k = 0$. For example, a similar derivation shows that where the signal has equal strength across all dimensions (i.e., $x_1 = \ldots = x_k = x$), confidence is minimized when $x = \frac{1}{2}\mu_\textrm{target}$ regardless of dimensionality:
\begin{align*}
    e^{\frac{1}{2}\mu_\textrm{target}^2} &= \frac{1}{k}\sum_{i=1}^k e^{x_i\mu_\textrm{target}} \\
    e^{\frac{1}{2}\mu_\textrm{target}^2} &= e^{x\mu_\textrm{target}} \\
    x = \frac{1}{2}\mu_\textrm{target}
\end{align*}

\section{Mechanistic Investigation of the Positive Evidence Bias in Neural Networks and Its Relation to Signal Detection Theory}\label{appendix:cnn}
Following the main finding that CNNs exhibit a Positive Evidence Bias (PEB) that increases with dimensionality, we sought to understand how this effect arises. In the original setting, we simultaneously increased the dimensionality of the convolutional encoder and the fully connected classifier heads. This made it unclear which network components were responsible for the increasing PEB. It also remained unclear to what extent the neural networks resemble the SDT model and which settings would enable the most meaningful comparison (see Discussion).

To resolve this, following a reviewer’s suggestion, we independently manipulated the dimensionality of the encoder and the classifier heads during training (with the test dimensionality equal to the head dimensionality during training). This allowed us to investigate whether the effect arises from (i) the encoder’s training dimensionality (representational richness), (ii) the classifier heads’ training and testing dimensionality (decoding of confidence from evidence), or (iii) an emergent property of their interaction.

Importantly, this is not the same as distinguishing between the training and testing dimensions, which could more closely resemble the SDT model. This is because, after training, the neural networks’ outputs become independent of one another, as the neural networks no longer learn. Thus, our results do not depend on whether the test set includes fewer or the same number of possible stimuli as the training set (we could, for example, remove some of the results afterwards, thereby creating an illusion of a dimensionality change). The only dimensionality discrepancy that can be meaningfully implemented is between encoder training and head training and testing.

\begin{figure}
    \centering
    \includegraphics[width=\linewidth]{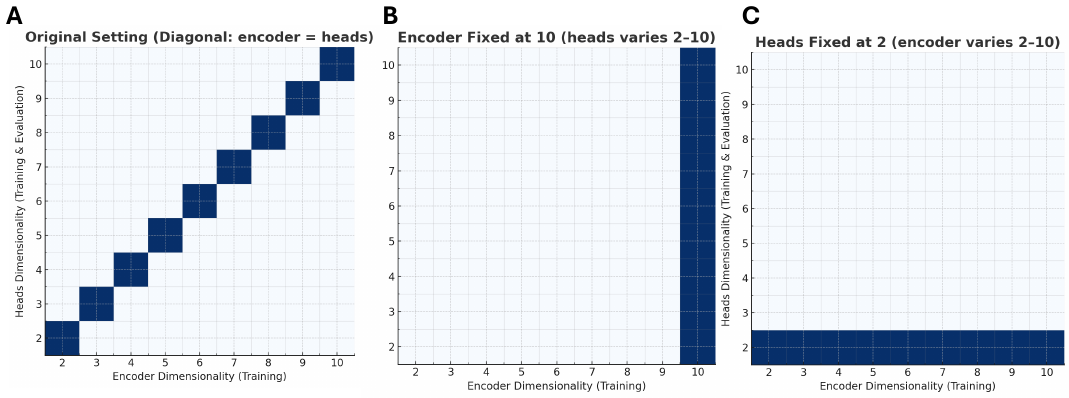}
    \caption{A) The original setting without freezing the encoder, in which the overall dimensionality of the network increases. B) Fixing the encoder dimension to check the influence of heads' dimensionality, C) Effect of the encoder dimensionality where the heads are fixed at dimensionality 2. The testing dimensionality is always equal to the heads’ training dimensionality.}
    \label{fig:training}
\end{figure}

To test these settings, we first trained CNNs on the full 10 dimensions. After training, we froze the encoder, keeping all its weights constant, and fine-tuned only the classifier heads on smaller subsets of the data. This manipulation allowed us to test whether the PEB depends on the dimensionality of the heads while the underlying representations remain fixed. In the complementary manipulation, we trained encoders on increasing numbers of dimensions (from 2 to 10), froze each encoder, and fine-tuned the heads for a fixed output dimensionality of 2. This reversed the logic of the first condition and tested whether the PEB depends instead on the encoder’s representational capacity (see Fig. \ref{fig:training}).

\subsection{Independent manipulation of encoder and head dimensionality}
\subsubsection{Freezing encoder trained on 10 dimensions, heads between 2-10 dimensions}
\begin{figure}
    \centering
    \includegraphics[width=.5\linewidth]{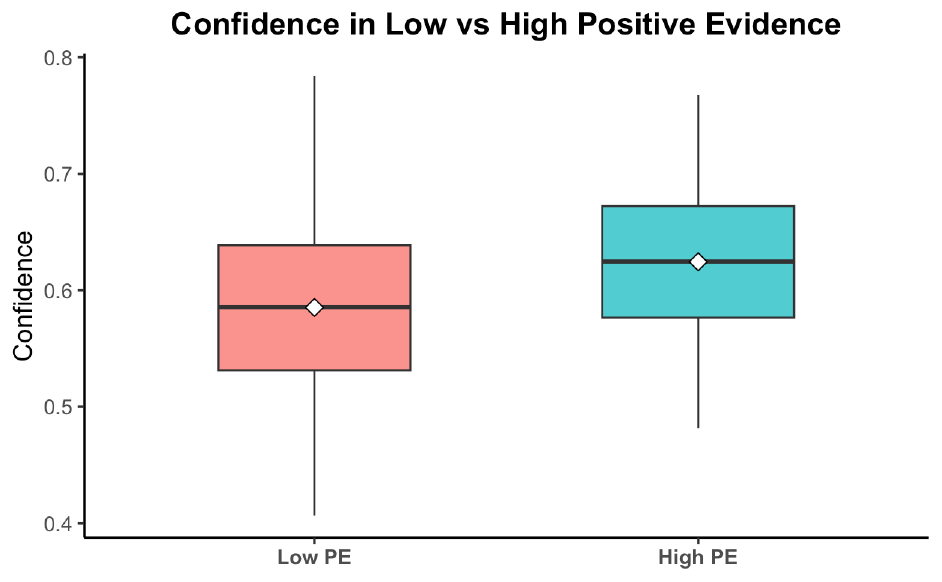}
    \caption{PEB with encoder dimensionality fixed at 10 dimensions, heads between 2-10 dimensions.}
    \label{fig:peb:heads:10:avg}
\end{figure}

We trained 225 convolutional neural networks, with 25 CNNs for each head dimensionality (ranging from 2 to 10), and evaluated at a varying achievable accuracy level (between 55-75\%), giving a total of 1125 CNNs output sets (5 accuracy levels x 9 head dimensionalities x 25 CNNs) for each Low and High Positive Evidence. Despite matching accuracy, confidence was significantly higher under High PE ($M=0.62$, $SD=0.06$) than Low PE ($M=0.59$, $SD=0.07$) (paired-$t(1124) = -33.82$, $p<.001$, 95\% CI = [-0.0414, -0.0369]) with a large effect size (Cohen’s $d=1.01$), demonstrating a very strong Positive Evidence Bias (PEB) (see Fig. \ref{fig:peb:heads:10:avg}).

\begin{figure}
    \centering
    \includegraphics[width=.5\linewidth]{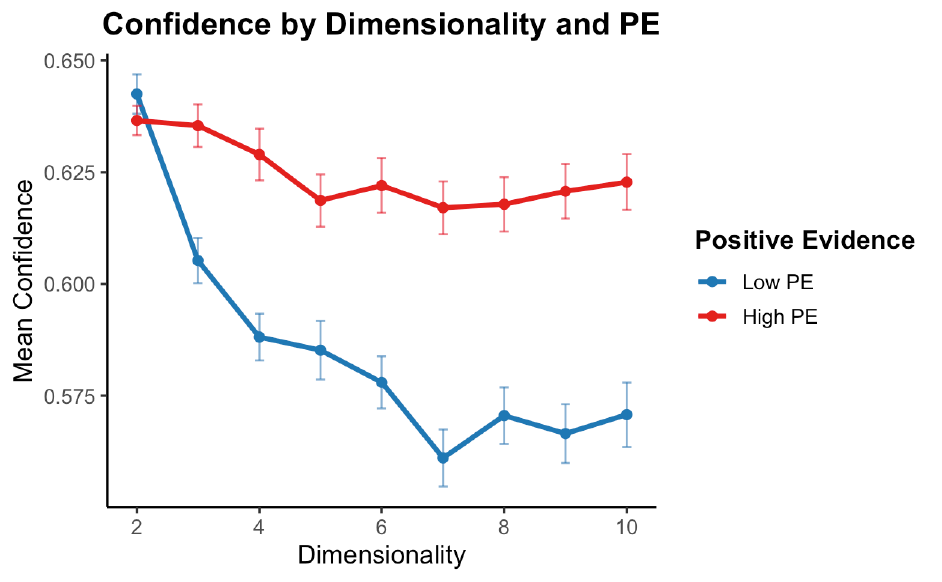}
    \caption{PEB increases with head dimensionality when encoder dimensionality is fixed to 10.}
    \label{fig:peb:heads:10}
\end{figure}

A mixed-effects model predicting confidence from z-scored head dimensionality and PE with random intercepts for network (confidence $\sim$ z(head dimensionality) × PE + (1$\vert$id)) revealed main effects of head dimensionality ($\beta=-0.02$, $t(1306)=-10.16$, $p<.001$) and change from low to High PE ($\beta=0.04$, $t(1125)=36.53$, $p<.001$), as well as a strong interaction ($\beta=0.01$, $t(1125)=13.63$, $p<.001$). This suggests that the PEB increased with head dimensionality (see Fig. \ref{fig:peb:heads:10}).

\subsubsection{Freezing encoder training at dimensionalities 2-10, with heads fixed at 2 dimensions}
\begin{figure}
    \centering
    \includegraphics[width=.5\linewidth]{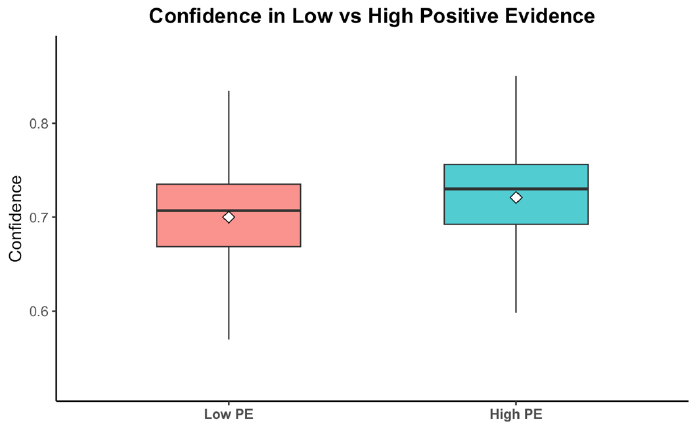}
    \caption{PEB for encoder dimensionality varying between 2 and 10 and head dimensionality fixed at 2 dimensions.}
    \label{fig:peb:encoder:2:avg}
\end{figure}

As in the previous manipulation, we trained 225 convolutional neural networks, with 25 CNNs for each encoder dimensionality (ranging from 2 to 10), and evaluated at across achievable accuracy levels (between 55-75\%), giving a total of 1125 CNNs output sets (5 accuracy levels x 9 encoder dimensionalities x 25 CNNs) for each Low and High Positive Evidence. Despite matching accuracy, confidence was significantly higher under High PE ($M=0.72$, $SD=0.05$) than Low PE ($M=0.7$, $SD=0.05$) (paired-$t(1124)=-27.59$, $p<.001$, 95\% CI=[-0.022, -0.019]) with a large effect size (Cohen’s $d=-0.82$), demonstrating large Positive Evidence Bias (PEB) effect (see Fig. \ref{fig:peb:encoder:2:avg}). 

\begin{figure}
    \centering
    \includegraphics[width=.5\linewidth]{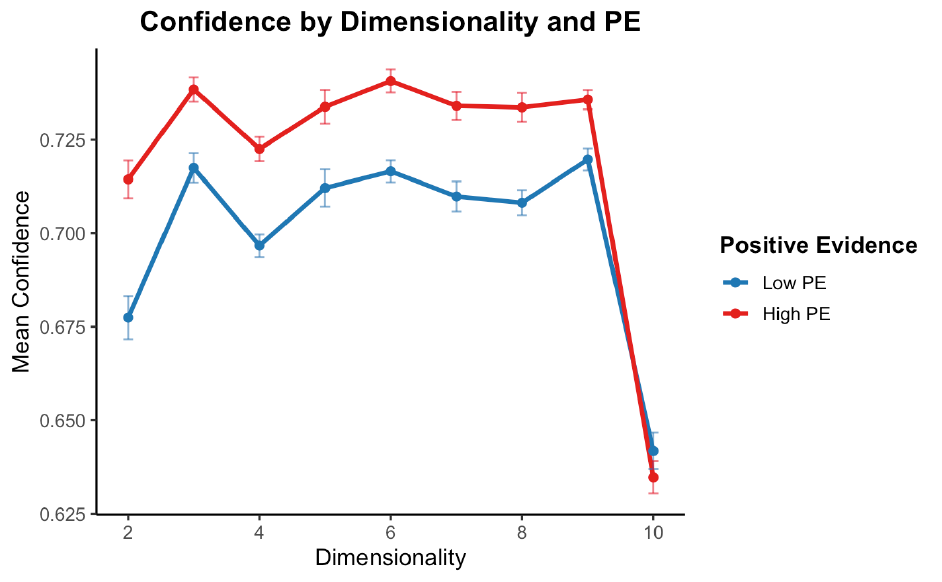}
    \caption{PEB does not increase with encoder dimensionality when head dimensionality is fixed to 2.}
    \label{fig:peb:encoder:2}
\end{figure}

A mixed-effects model predicting confidence from z-scored encoder dimensionality and PE with random intercepts for NN (confidence $\sim$ z(encoder dimensionality) × PE + (1$\vert$ id)) revealed main effects of encoder dimensionality ($\beta=-0.005$, $t(1256)=-3.24$, $p=.001$) and change from low to High PE ($\beta=0.02$, $t(1125)=29.13$, $p<.001$) and an interaction ($\beta=-0.008$, $t(1125)=-11.33$, $p<.001$). Thus, the PEB was slightly decreasing with encoder dimensionality (see Fig. \ref{fig:peb:encoder:2}). We also found an unexpected pattern: confidence was lower overall and the PEB was smaller when the encoder dimensionality was 10. It is unclear why exactly the results for this dimensionality differ from the lower encoder dimensionalities or why the transition between 9 and 10 dimensions is so abrupt. However, this change could still not explain our main finding, i.e., that the PEB \emph{increases} with dimensionality.

\subsubsection{Conclusion}
Across all configurations, the PEB was consistently observed: confidence was higher in the high-positive-evidence than in the low-positive-evidence condition, under matched accuracy. However, the strength of the effect increased only when the dimensionality of the classifier heads increased while the dimensionality of the encoder was fixed. When the encoder dimensionality increased but the head dimensionality remained fixed, the PEB did not strengthen. These results suggest that the rise of the PEB with overall dimensionality originates from the classifier heads rather than from the encoder. Increasing the capacity of the heads that transform internal representations into confidence estimates amplified the bias, even when the encoder's representational space remained unchanged. This pattern indicates that the mapping from evidence to confidence, rather than evidence encoding \emph{per se}, is the key source of the bias within CNNs.

\subsection{Mid-value cases---how generalisable is the effect?}
\begin{figure}
    \centering
    \includegraphics[width=.33\linewidth]{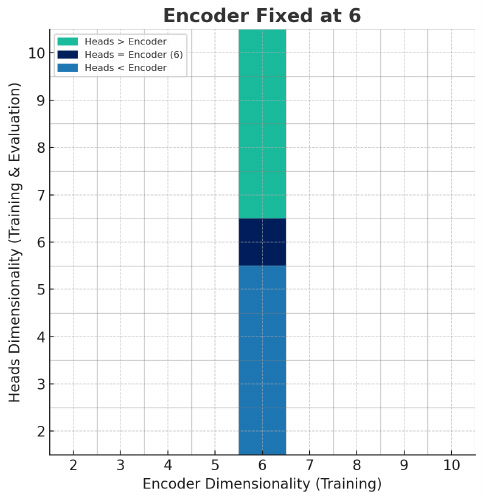}
    \caption{Checking whether the effect generalises to lower dimensions (see Fig 1). Taking a mid-value 6, we additionally have a balanced number of CNNs with dim(encoder) $>$ dim(heads) and dim(encoder) $<$ dim(heads).}
    \label{fig:training:heads}
\end{figure}

As shown in Figure \ref{fig:training}, we only investigated the edge cases of the encoder's maximum dimension or the heads' minimum dimension. This was to maximise the sample size while avoiding a situation in which the heads are trained and tested on stimuli the encoder has not seen. To further test the generalisability of this effect, we explored intermediate cases in which the encoder or heads were frozen at a mid-range dimensionality (see Figs. \ref{fig:training:heads}, \ref{fig:training:encoder}) while the head dimensionality varied across the full range. These configurations balanced cases where the encoder had higher, equal, or lower dimensionality than the heads. Importantly, lowering the encoder dimensionality or increasing the head dimensionality is inherently linked to problems with reaching high accuracy (e.g., for the lowest encoder dimensionalities, there is no such high signal value to achieve 0.75 accuracy). Therefore, not all cases can be checked, at least not for all levels of accuracy. In particular, we cannot check low encoder or high head dimensionalities at high accuracies. 

We trained a total of 900 CNNs, with 25 CNNs for each head dimensionality (ranging from 2 to 10), with encoder's dimensionality fixed at the mid-level (6), and evaluated across 4 achievable accuracy levels (55-70\%) for Low and High Positive Evidence. The paired t-test showed a significant PEB ($t(899)=-42.62, p<.001$), with a large effect size (Cohen's $d = 0.63$).

\begin{figure}
    \centering
    \includegraphics[width=\linewidth]{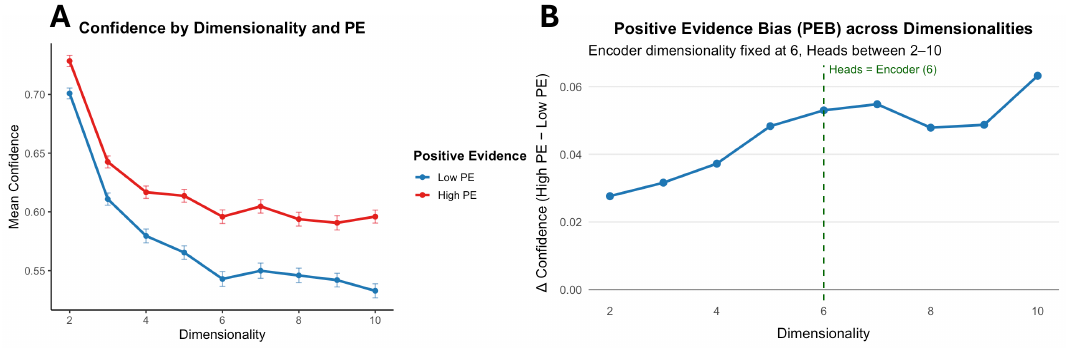}
    \caption{PEB increases with head dimensionality when the encoder dimensionality is frozen at 6 and the head dimensionality ranged from 2 to 10. B) Depicts the effect, highlighting the moment when head dimensionality crosses encoder dimensionality.}
    \label{fig:peb:heads:6}
\end{figure}

\begin{figure}
    \centering
    \includegraphics[width=.33\linewidth]{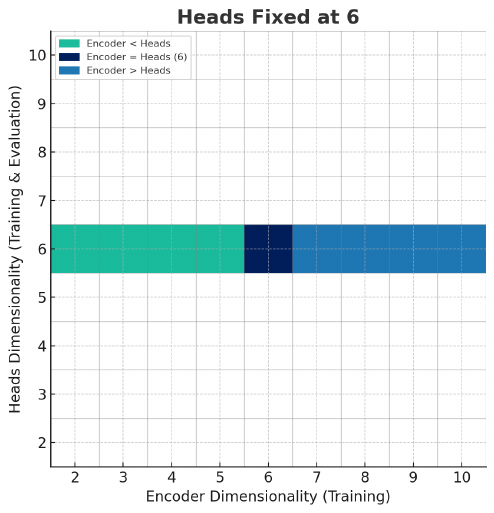}
    \caption{Checking the effect of encoder dimensionality with head dimensionality fixed at mid-value 6.}
    \label{fig:training:encoder}
\end{figure}

A mixed-effects model predicting confidence from z-scored head dimensionality and Positive Evidence (PE), with random intercepts for network ID (confidence $\sim$ z(head dimensionality) × PE + (1$\vert$id)), revealed significant main effects of head dimensionality ($\beta = -0.041$, $t(1015) = -19.82$, $p < .001$) and PE ($\beta = 0.046$, $t(900) = 44.63$, $p < .001$), as well as a significant interaction ($\beta = 0.010$, $t(900) = 9.27$, $p < .001$). Confidence decreased with increasing head dimensionality and PE, but PEB was stronger at higher dimensionality, as with the encoder frozen at 10 dimensions. Note that although a logarithmic model would fit the data better (see Fig. \ref{fig:peb:heads:6}), we use the less powerful linear model for the sake of comparison with other (non-logarithmic) cases.

The resulting pattern again supports the conclusion that a PEB arises from the classifier heads: confidence increased strongly for high-evidence trials when head dimensionality increased, regardless of whether the encoder was richer or poorer in features. Similar results were later replicated on dimension five and, to some extent, even in the smallest possible encoder dimension (2). The effect in the latter could not be fully investigated because the value was so small that the neural networks failed to achieve meaningful performances for head dimensions above 4. This is likely because if the encoder is trained on a very small subset of digits, its representation is not rich enough to be used when the head's dimensionality is high. Nevertheless, the achievable results did not deviate from the expected pattern, suggesting that the head’s dimensionality indeed drives the PEB in the neural networks across accuracy levels and encoder dimensionalities. 

Finally, the effect was checked manipulating the encoder dimensionality with head dimensionality fixed to 6, as shown in Figure \ref{fig:training:encoder}.
A total of 675 CNNs were trained, with 25 networks for each encoder dimensionality (ranging between 2 and 10), with heads dimensionality fixed at 6; and evaluated across 3 achievable accuracy levels (55-65\%). There was a significant PEB (paired-$t(674)=-29.75, p<.001$) with a large effect size (Cohen's $d=0.87$).

\begin{figure}
    \centering
    \includegraphics[width=\linewidth]{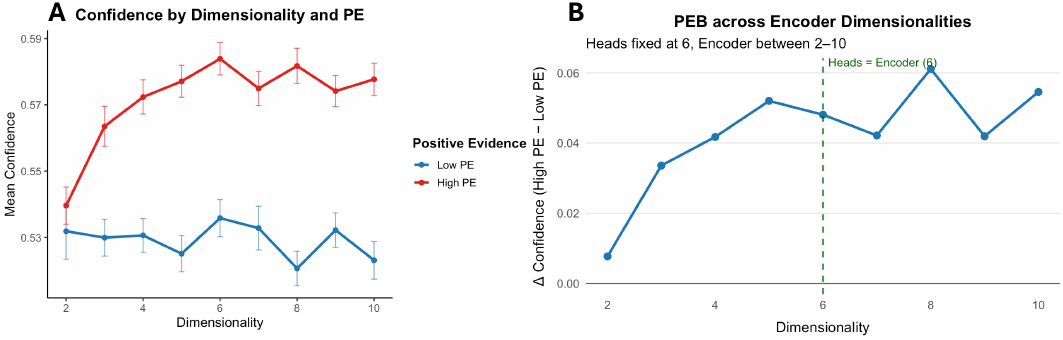}
    \caption{Although the confidence pattern differs for fixed head dimensionality, the PEB increases with the encoder's dimensionality when the head dimensionality is fixed to 6. B) Shows the effect by highlighting the moment when encoder dimensionality crosses head dimensionality.}
    \label{fig:peb:encoder:6}
\end{figure}

A mixed-effects model predicting confidence from z-scored encoder dimensionality and Positive Evidence (PE), with random intercepts for network ID (confidence $\sim$ scale(encoder dimensionality) × PE + (1$\vert$id)), showed that confidence did not significantly vary with encoder dimensionality ($\beta = -0.0017$, $t(879.6) = -0.93$, $p = .35$), but increased significantly between the low and high PE conditions ($\beta = 0.0425$, $t(675) = 31.01$, $p < .001$). Critically, the interaction between encoder dimensionality and PE was significantly positive ($\beta = 0.0104$, $t(675) = 7.56$, $p < .001$), indicating that the Positive Evidence Bias (PEB) increased with the encoder’s dimensionality when the heads’ dimensionality was fixed at 6 (Figure \ref{fig:peb:encoder:6}). In contrast to the extreme edge cases, this shows that for the mid-value head dimensionality, the PEB increases with both head dimensionality (with encoder dimensionality fixed to 6) and encoder dimensionality (with head dimensionality fixed to 6). This finding indicates that, for some values, the effect of dimensionality is visible independently in both the encoder and the heads. Note that the accuracy here is limited because it was not possible to achieve 70\% accuracy with the low encoder dimensions. However, when including models that were able to achieve 70\% accuracy, the effect is the same (with an additionally significant main effect of dimensionality).

\subsection{Discussion}
The present analyses aimed to isolate which components of convolutional neural networks (CNNs) give rise to the Positive Evidence Bias (PEB) and its increase with dimensionality. The PEB was observed across all manipulations and, crucially, its magnitude varied across dimensions, increasing primarily as the dimensionality of the fully connected heads increased, even when the classifier was frozen at different dimensionalities. However, the PEB also increased when the heads were fixed on mid-values, and the encoder’s dimensionality increased. This pattern might indicate that the bias can arise from both the mapping between internal representations and confidence outputs, as well as from the independent encoding of sensory evidence. The transformation of internal evidence into a confidence judgment appears to be a robust locus of the PEB within CNNs, but its dependence on the heads is not absolute, as a dimensionality-dependent PEB can also be observed independently in the encoder.

Although these results are informative for understanding how the PEB manifests within CNNs, they offer limited traction for mechanistic comparison with the Signal Detection Theory (SDT) model. The two frameworks share the same computational-level goal---to optimise decision and confidence under uncertainty---but they differ profoundly at the algorithmic level. CNNs learn mappings between high-dimensional representations and decision outputs via gradient-based optimisation, whereas SDT models confidence as a probabilistic inference over noisy evidence. Interpreting the encoder–head discrepancy as analogous to evidence versus evaluation dimensionality may loosely reflect Bayesian accounts of the PEB, but importantly, CNNs cannot emulate a ``training versus testing'' dimensionality discrepancy: once trained, each test trial is independent. Both the encoder and the heads are trained jointly, making evidence-space vs. evaluation-space analogies ambiguous. The correspondence between CNNs and SDT is therefore more straightforward at the computational level but ambiguous at the algorithmic level.

Additionally, a universal limitation of interpreting the CNN results is linked to the multiple researchers’ degrees of freedom, including the choice of architecture, training regime, dataset composition, and regularisation---all of which could, in principle, affect outcomes. However, the robustness of the PEB across all our manipulations demonstrates that this bias is not an artefact of selective model tuning. Additionally, no arbitrary choices were made during the process as the architecture used the CNN model from \citet{Webb2023}, which already exhibited the PEB, and simply manipulated its dimensionality to test whether the bias scales with representational complexity. We further confirmed that the same pattern persists across multiple accuracy levels, extending beyond the single accuracy tested in the original work. The replication of the dimensionality-related PEB under these constraints indicates that it is a genuine property of these networks rather than a product of arbitrary design choices.

Taken together, these analyses indicate that the PEB in these neural networks likely arises as a by-product of how these multidimensional, high-capacity systems optimise the transformation from evidence to confidence under uncertainty. The similarity to human PEB therefore resides at the computational level: both systems produce overconfidence when operating in high-dimensional representational spaces. However, we do not claim that any particular part of the neural network directly corresponds to a component of the SDT model. The analogy between CNNs and SDT is thus informative but bounded---useful for illustrating how confidence biases can emerge from optimisation in complex representational spaces, but not for inferring mechanistic equivalence between the two frameworks.

\section{Additional control analyses}
A reviewer pointed out one potential confound in our neural network simulations: namely, that when manipulating dimensionality, we fixed training accuracy at a value from 55\% for all dimensionalities. This approach raises two concerns. First, our results might be sensitive to the particular value of absolute accuracy we chose. Second, our results might be confounded by relative (not absolute) accuracy. Here we report two sets of simulations addressing these two concerns.

\subsection{Simulations controlling for absolute accuracy}

First, we investigated whether our results were robust when training networks to a different fixed level of performance. For each fixed-accuracy dataset, we fit a mixed-effects model predicting confidence from PE condition, z-scored log dimensionality, and their interaction, with a random intercept for network identity:
\[
\text{confidence} \sim z(\log(\text{dimension})) \times \text{PE} + (1|\text{id}).
\]

\begin{figure}
    \centering
    \includegraphics[width=0.75\linewidth]{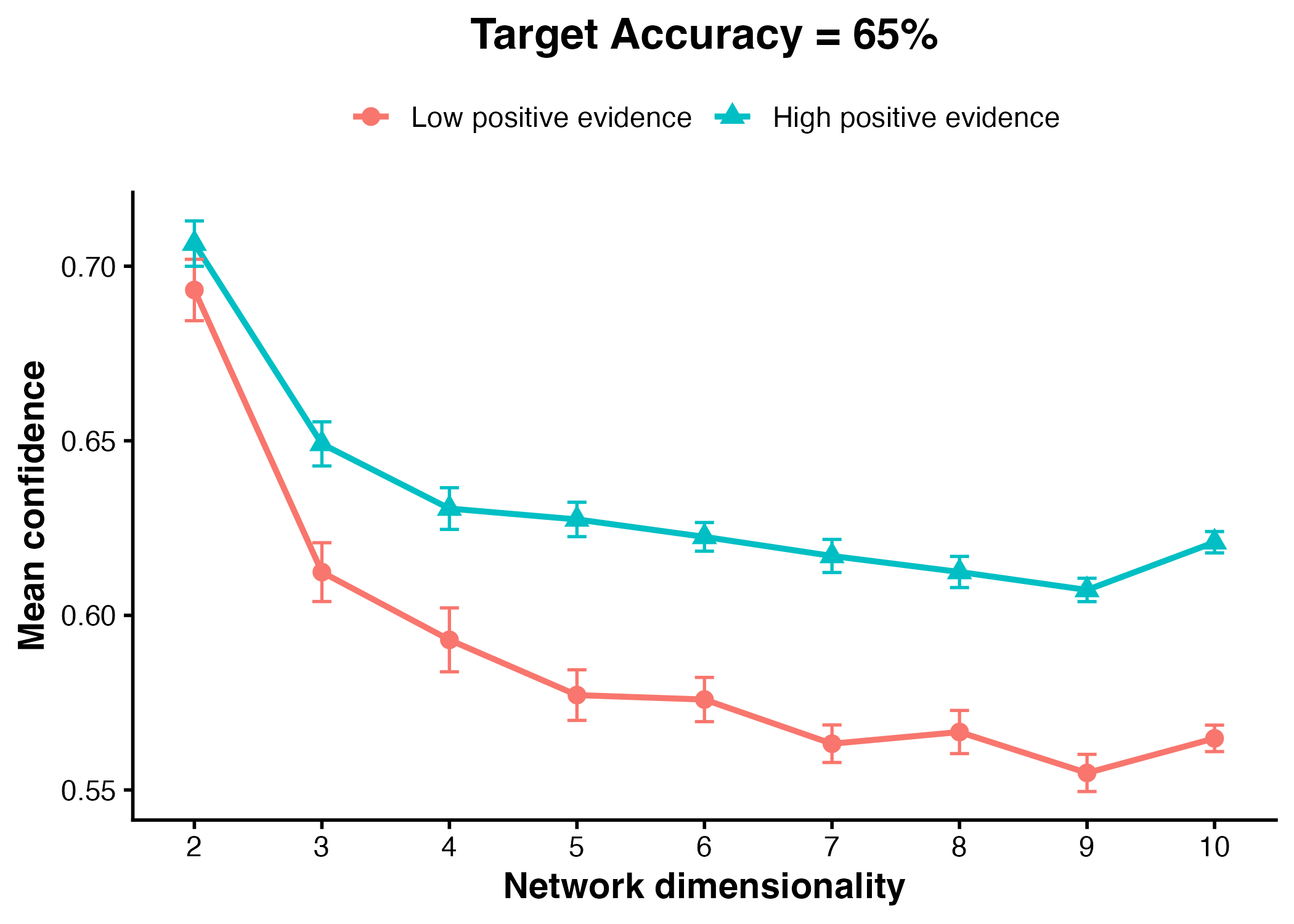}
    \caption{Mean confidence as a function of positive evidence and dimensionality with training accuracy set to 65\%.}
    \label{fig:peb_65}
\end{figure}

At 65\% accuracy, the dimensionality $\times$ PE interaction was positive and significant ($\beta = 0.011$, $SE = 0.002$, $t = 6.23$, $p < .001$, Figure \ref{fig:peb_65}). Thus, even when networks were trained to a higher fixed accuracy level than in the original simulations, the PEB continued to increase with dimensionality. This indicates that the dimensionality effect is not specific to the original 55\% training-accuracy setting.

\begin{figure}
    \centering
    \includegraphics[width=0.75\linewidth]{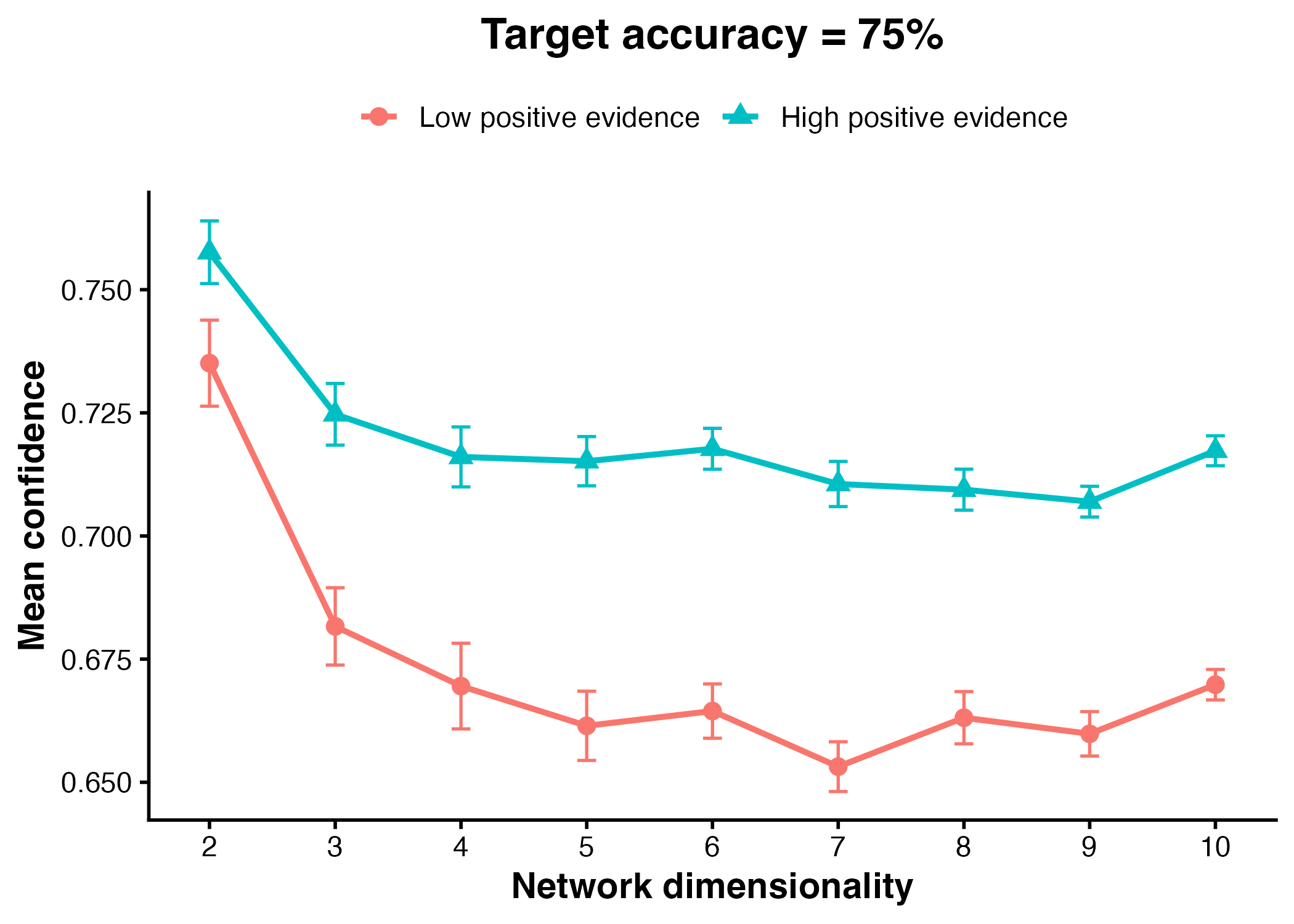}
    \caption{Mean confidence as a function of positive evidence and dimensionality with training accuracy set to 75\%.}
    \label{fig:peb_75}
\end{figure}

We then repeated the same analysis with networks trained to 75\% accuracy (Figure \ref{fig:peb_75}). The dimensionality $\times$ PE interaction again remained positive and significant ($\beta = 0.006$, $SE = 0.002$, $t = 4.14$, $p < .001$), with the PEB still increasing with dimensionality.

To further assess the role of accuracy in driving the PEB, we fit a mixed-effects model using the full set of CNN outputs across all accuracy levels (0.55–0.80; PEB $\sim$ log(dimension)*accuracy). The model revealed that PEB increased with accuracy ($\beta = 0.13$, $SE = 0.03$, $p < .001$) and dimensionality ($\beta = 0.05$, $SE = 0.01$, $p < .001$). There was a negative interaction ($\beta = –0.06$, $SE = 0.017$, $p < .001$), indicating that dimensionality contributes more strongly to the PEB when accuracy is lower. Together, these results establish that while the strength of the effect of dimensionality on the PEB varies between accuracy levels, it generalises across the entire achievable accuracy range.

\subsection{Simulations controlling for relative accuracy}
\begin{figure}
    \centering
    \includegraphics[width=0.5\linewidth]{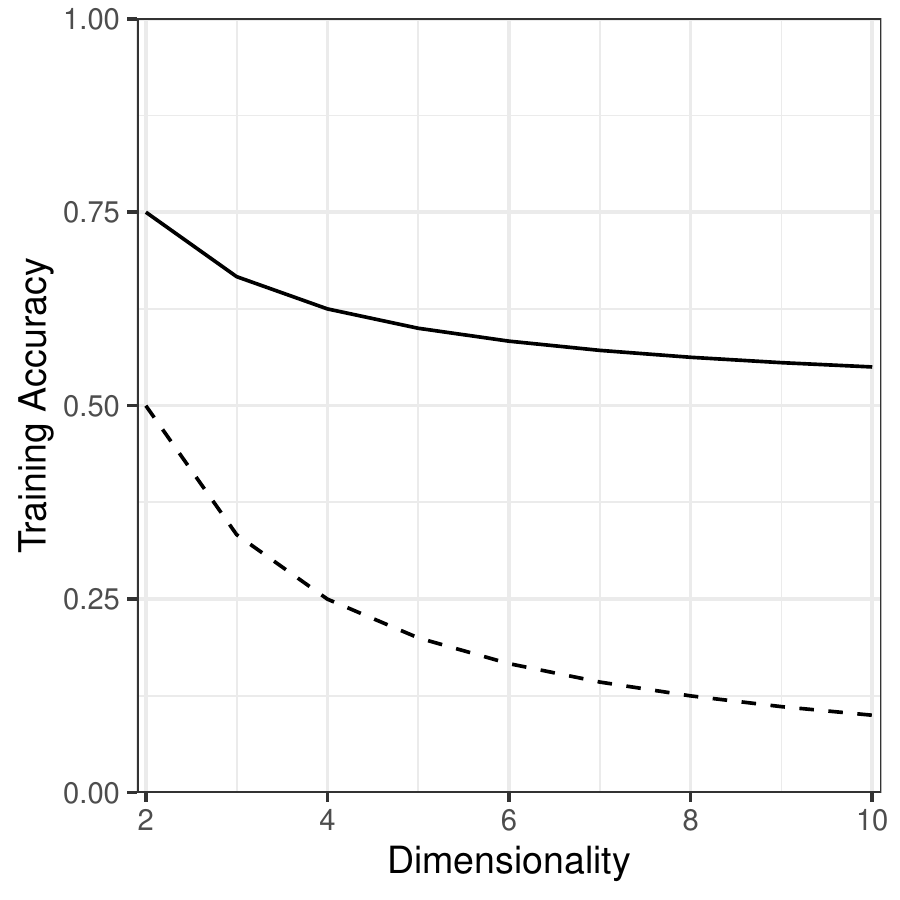}
    \caption{Training accuracy used for each dimensionality (solid line) against chance performance (dashed line).}
    \label{fig:relative_accuracy}
\end{figure}

The second concern is that while our networks achieved the same absolute accuracy, this same level of accuracy is at near-chance for the network with $k=2$ but halfway between chance and perfect performance when $k=10$ (similarly, an accuracy of 75\% is halfway between chance and perfect performance for $k=2$ but near ceiling for $k=10$). This leaves open the possibility that differences in the PEB as a function of dimensionality could arise simply if the networks trained to more sensitive ranges of performance exhibited larger biases. To control for this, we ran another set of simulations training each network to its most sensitive accuracy level (i.e., one that is halfway between chance and perfect performance; see Figure \ref{fig:relative_accuracy}). Note that while controlling for relative performance, this analysis necessarily confounds results by absolute performance: one cannot control for both simultaneously.

\begin{figure}
    \centering
    \includegraphics[width=0.75\linewidth]{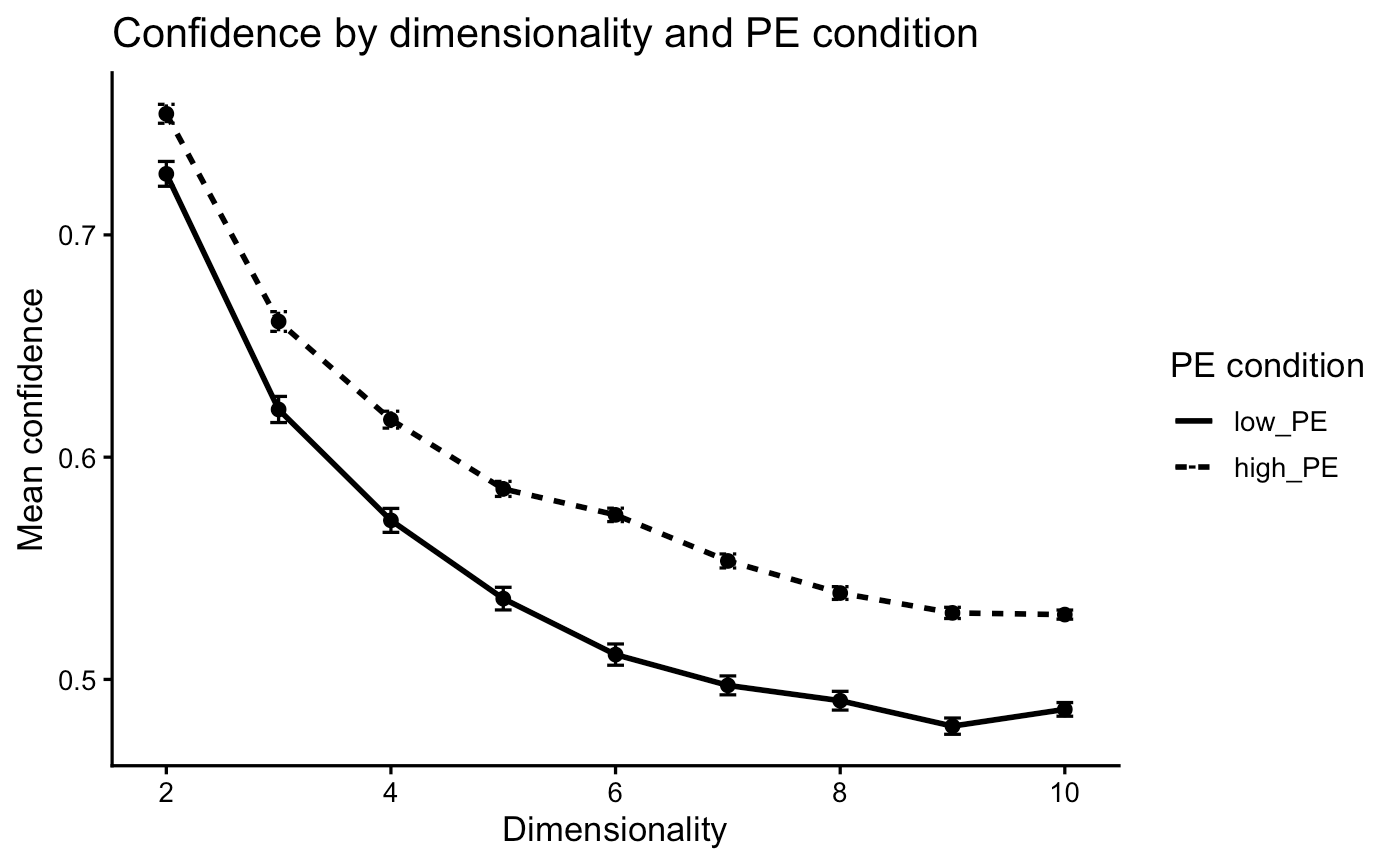}
    \caption{Mean confidence as a function of positive evidence and dimensionality, controlling for relative (as opposed to absolute) accuracy across dimensionalities.}
    \label{fig:peb_relative}
\end{figure}

We repeated the confidence analysis on CNNs evaluated at the midpoint sensitivity level using 100 neural networks, increasing the sample size to improve statistical power. We fit a linear mixed-effects model predicting confidence from PE condition, scaled dimensionality, and their interaction, with random intercepts for source file and run. The model revealed a significant main effect of PE condition, $\beta = 0.047$, $SE=0.001$, $t(887)=35.77$, $p<.001$, such that confidence was higher in the high-PE than low-PE condition. There was also a significant main effect of dimensionality, $\beta=-0.068$, $SE=0.01$, $t(7.04)=-5.56$, $p<.001$, indicating lower overall confidence at higher dimensionalities. Importantly, the PE Condition × Dimensionality interaction was significant, $\beta=0.0046$, $SE=0.001$, $t(887)=3.48$, $p<.001$, indicating that the positive evidence bias increased with dimensionality (see Figure \ref{fig:peb_relative}). This suggests that differences in the accuracy level associated with the midpoint sensitivity criterion were not a critical confound driving the original pattern of results.

\end{document}